\documentclass[12pt]{article}
\usepackage{eurosym}
\usepackage{amsfonts}
\usepackage{amssymb,amsmath}
\usepackage{graphicx}
\usepackage{young}
\usepackage{youngtab}
\usepackage{color}
\usepackage{subcaption}
\usepackage{ytableau}

\numberwithin{equation}{section}

\newcommand{\be}{\begin{equation}}
\newcommand{\ee}{\end{equation}}
\newcommand{\bea}{\begin{eqnarray}}
\newcommand{\eea}{\end{eqnarray}}
\newcommand{\beas}{\begin{eqnarray*}}
\newcommand{\eeas}{\end{eqnarray*}}
\newcommand{\ba}{\begin{array}}
\newcommand{\ea}{\end{array}}

\newcommand{\nn}{\nonumber}

\newcommand{\nbox}{{\,\lower0.9pt\vbox{\hrule \hbox{\vrule height 0.2 cm \hskip 0.19 cm \vrule height 0.2 cm}\hrule}\,}}
\newcommand{\Tr}{\ {\rm Tr}\ }

\def\href#1#2{#2}
\begin{document}

\begin{titlepage}
\hfill
\vbox{
    \halign{#\hfil         \cr
           } 
      }  

\hbox to \hsize{{}\hss \vtop{ \hbox{}

}}


\vspace*{20mm}
\begin{center}

{\large \textbf{Counting paths with Schur transitions}}

\vspace{ 8mm}

{\normalsize {Pablo D\'iaz${}^{1}$, Garreth Kemp${}^{2}$, and Alvaro V\'eliz-Osorio${}^{3,4}$}  }

{\normalsize \vspace{10mm} }

{\small \emph{${}^1$\textit{Department of Physics and Astronomy, University of Lethbridge, \\
Lethbridge,
Alberta, T1K 3M4, Canada
}} }

{\normalsize \vspace{0.4cm} }

{\small \emph{$^2$\textit{ Department of Physics, University of Johannesburg,\\ P.O. Box 524, Auckland Park 2006, South Africa
}} }

{\normalsize \vspace{0.2cm} }

{\small \emph{$^3$\textit{Mandelstam Institute for Theoretical Physics, University of the Witwatersrand,\\ WITS 2050, Johannesburg, South Africa
}} }

{\normalsize \vspace{0.4cm} }

{\small \emph{$^4$\textit{ School of Physics and Astronomy, Queen Mary, University of London,\\ Mile End Road, London E1 4NS, United Kingdom
}} }

{\normalsize \vspace{0.4cm} }
%
\end{center}

\begin{abstract}

In this work we explore the structure of the branching graph of the unitary group using Schur transitions. 
We find that these transitions suggest a new combinatorial expression for counting paths in the branching graph. This formula, which is valid for any rank of the unitary group, reproduces known asymptotic results. We proceed to establish the general validity of this expression by a formal proof. The form of this equation strongly hints towards a quantum generalization. Thus, we introduce a notion of quantum relative dimension and subject it to the appropriate consistency tests. This new quantity finds its natural environment in the context of RCFTs and fractional statistics; where the already established notion of quantum dimension has proven to be of great physical importance.

\end{abstract}

\end{titlepage}

\tableofcontents
\newpage


\section{Introduction}

\label{sec: introduction}

A great deal of information about physical systems can be encoded in the group theoretical structure of its symmetries. Often, the states of a theory can be labeled by irreducible representations (irreps) of its symmetry group. Moreover, irreps of a group can be decomposed into irreps of its subgroups. The \emph{branching rules} tell you the way this decomposition is carried out and the process can be depicted in terms of a so-called branching graph. Branching rules have relevant physical applications; for instance, in phenomena that involve symmetry breaking such as the Zeeman effect. In this article we explore some interesting properties of branching graphs.

The relationship between the branching graph of the unitary groups and the branching graph of the symmetric groups was extensively studied by Borodin and Olshanski (BO) in \cite{BO}. Their results can be seen as an extension of the celebrated Schur-Weyl duality. Inspired by the success of the latter in encoding some features of the AdS/CFT correspondence \cite{Ramgoolam:2008yr}, one of the identities found in \cite{BO} involving both branching graphs was studied from a holographic perspective in \cite{Diaz:2015tda}. In that work it was found that the data naturally associated with one side of the BO identity, namely the one linked to symmetric groups, is exactly reproduced by CFT three-point functions related to special backgrounds. From this observation we infer that the BO identity is endowed with physical content.

The BO identity is valid near the boundary of the unitary group's branching graph, that is, when the rank ($N$) tends to infinity. However, in this article we explore the structure of this graph when $N$ is finite. Surprisingly, computing the same type of three-point function as mentioned above, this time at finite $N$, reveals the exact structure of the unitary branching graph. This structure  is encoded in a set of probabilities naturally associated with the paths of the graph. The probabilities involve the concept of {\it relative dimension} of two irreps in the graph, which counts the number of paths which join the given irreps. Relative dimensions are hard to compute except for simple cases. However, by approaching them through our three-point function computations we are able to give a compact combinatorial form for the relative dimension, see Eq. (\ref{conjecture}). This formula is surprisingly simple (and easy to prove!) and, as far as we are aware, it has not appeared in the literature before. Moreover, the BO identity follows as a simple corollary of this formula.

Another appealing feature of our formula for the relative dimension is that it admits a natural generalization to the realm of affine Lie algebras. These algebras are of central importance in the study of rational conformal field theories (RCFT) \cite{Moore:1989vd} and fractional statistics \cite{Wen:1990se,Moore:1991ks}. The Hilbert spaces of this kind of models can be arranged into a finite number of representations (families) of their underlying affine Lie algebras. The \emph{size} of each of these families in the Hilbert space is captured by a quantity called the \emph{quantum dimension}; which reduces, in the classical limit, to the dimension of the representation itself. The point is, that all the objects appearing in our formula for the relative dimension naturally admit quantum generalizations. We are thus lead to propose a definition for a \emph{quantum relative dimension} in Eq. (\ref{ReQD}), whose consistency we show. Quantum dimensions have some interesting physical interpretations in terms of quantum entanglement \cite{Kitaev:2005dm, He:2014mwa, Caputa:2015tua} and quantum chaos \cite{Caputa:2016tgt, Gu:2016hoy}. We hope that this notion of  quantum relative dimension might find similar applications in the future.

This paper is organized as follows: Section \ref{sec: GT and Y graphs} presents the relevant branching graphs and introduces the key concept of relative dimension. In section \ref{sec: multi-graviton transitions}, we compute certain three-point functions for Schur states at finite $N$ and show how they secretly encode a formula for the relative dimension of the unitary graph. Then, in section \ref{proof} we give a proof for this formula. Finally, section \ref{QRD} introduces the concept of quantum relative dimension. We place a number of examples and technical details in the appendices. 

\section{GT graph, Young graph and BO identity}

\label{sec: GT and Y graphs}

 Recently, a tantalizing relationship between two representation theoretical graphs has been uncovered by Borodin and Olshanski \cite{BO}. On one side, we have the Young graph ($\mathbb{Y}$) describing the branching of symmetric groups, while on the other we have the Gelfand-Tsetlin graph ($\mathbb{GT}$) depicting that of unitary groups. In some sense, their result can be viewed as an extension of the celebrated Schur-Weyl duality. In this section we introduce these two graphs and present the duality relating them.

\subsection{The Young graph}

First, we consider the Young graph, whose vertices are given by Young diagrams. The graph is leveled by the number of boxes in each diagram. Clearly, this graph is infinite since it is possible keep climbing by adding boxes indefinitely. Vertices in $\mathbb{Y}$ are connected if and only if their corresponding Young diagrams can be obtained from each other by adding or removing a single box. Recalling that Young diagrams with $n$ boxes label irreducible representations (irreps) of the symmetric group $S_n$, it is possible to give a group-theoretic interpretation to $\mathbb{Y}$; namely, the Young graph represents how irreps of $S_n$ are subduced by irreps of $S_{n+1}$ for each level $n$. Hereafter, we will reserve the letters $m$ and $n$ to label the levels on this graph, while the letters $\mu$ and $\nu$ will stand for Young diagrams.

 Notice that from any given vertex $\mu  \in \mathbb{Y}$
it is possible to follow at least one path downwards all the way to the bottom.
 Each of those paths corresponds to a way of decomposing the Young diagram $\mu$ one box at a time. 
In group theory terminology, each of these paths corresponds to a list of linked irreps associated with the chain of embeddings:
\begin{equation}
S_{n}\supset S_{n-1}\supset \cdots \supset S_{1}.
\end{equation}%
\newline
\newline
\newline
\newline
\newline
\setlength{\unitlength}{2.7cm} {\small
\begin{picture}(3,1)\label{1}
\put(0,1.1){$n=3$}
\put(1.5,1){$\begin{Young}
&\cr
\cr
\end{Young}$}
\put(.7,1){$\begin{Young}
\cr
\cr
\cr
\end{Young}$}
\put(2.5,1){$\begin{Young}
&&\cr
\end{Young}$}
\put(0,.5){$n=2$}
\put(1,.5){$\begin{Young}
\cr
\cr
\end{Young}$}
\put(2,.5){$\begin{Young}
&\cr
\end{Young}$}
\put(.95,.82){\line(-1,1){.14}}
\put(2.3,.68){\line(1,1){.27}}
\put(1.5,.95){\line(-1,-1){.3}}
\put(1.7,1){\line(1,-1){.32}}
\put(0,-.2){$n=1$}
\put(1.5,-.2){$\begin{Young}
\cr
\end{Young}$}
\put(1.55,0){\line(-1,1){.4}}
\put(1.65,0){\line(1,1){.45}}
\end{picture}} \newline
\newline
\newline
As a matter of fact, the number of paths descending from $\mu$ matches the dimension of the irrep $\mu$; with each path corresponding to a state of the irrep. The dimension of $\mu$ can be computed by means of the so-called hook length formula as follows.
To start, if $(i,j)$  is a cell in $\mu$, then its \textit{hook} is the set
\begin{equation}\label{hook length}
\text{H}_\mu(i,j)=\{(a,b)\in \mu| a=i, \, b\geq j\}\cup \{(a,b)\in \mu| b=j, \, a\geq i\},
\end{equation}
and its \textit{hook length} is given by $h_\mu(i,j)\equiv |\text{H}_\mu(i,j)|$.
The hook length of the diagram $\mu$ is simply \footnote{Frequently, the notation $\text{Hooks}_\mu$ is used for $H_\mu$, and we choose the latter to avoid long expressions in the following sections.}
\begin{equation}\label{hook length}
 H_\mu= \prod_{(i,j)\in\mu} h_\mu(i,j)\,.
\end{equation}
Using this quantity, the dimension of the irrep $\mu$ is given by 
\begin{equation}\label{dim hook}
\text{dim}_{\mu }=\frac{m!}{H_\mu}.
\end{equation}%
Another notion that will be central to our discussion is that of the relative dimension $\text{dim}(\mu ,\nu )$ (with $m\geq n$ ) of two irreps $\mu$ and $\nu$, which is
\begin{equation}
\text{dim}(\mu,\nu)=(\text{\# paths from }\mu \text{ to }\nu)
\end{equation}


Now, we add an extra layer of structure to this graph and we do so in a more general setting. Let $\mathbb{G}$ be an arbitrary leveled graph. Given a pair of vertices $\mu, \nu\in \mathbb{G}$
such that $\mu$ is at a higher level than $\nu$ we define the quantity
\begin{equation}\label{proba}
{}^{\mathbb{G}}\Lambda _{n}^{m} (\mu ,\nu )\equiv\left(\frac{\text{\# paths from $\nu$ to the ground floor}}{\text{\# paths from $\mu$ to the ground floor}}\right)\times\left(\text{\# paths from $\mu$ to $\nu$}\right),
\end{equation}%
where $m$ and $n$ are the respective levels of the vertices. Observe that Eq.\,\eqref{proba} satisfies
\begin{equation}\label{nor}
\sum_{\nu}{}^{\mathbb{G}}\Lambda _{n}^{m}(\mu ,\nu )=1,
\end{equation}%
where the sum runs over all the vertices $\nu$ at level $n$. Thus, for a fixed $\mu\in \mathbb{G}$, the quantity \eqref{proba} furnishes a probability distribution on each level $n<m$  in the graph. Moreover, these distributions satisfy the compatibility condition
\begin{equation}\label{compa}
\sum_{\nu'}{}^{\mathbb{G}}\Lambda _{n^{\prime
}}^{m}(\mu ,\nu^{\prime }){}^{\mathbb{G}}\Lambda _{n}^{n^{\prime }}(\nu
^{\prime },\nu )={}^{\mathbb{G}}\Lambda _{n}^{m}(\mu,\nu )\,
\end{equation}
for any intermediate level, i.e. $n<n'<m$.

The above construction is valid for any leveled graph, $\mathbb{Y}$ for instance. In terms of the irrep's dimensions, Eq.\,\eqref{proba}  can be expressed as
\begin{equation}\label{prob young}
{}^{\mathbb{Y}}\Lambda _{n}^{m}(\mu ,\nu )= \frac{\text{dim}_{\nu }%
}{\text{dim}_{\mu }} \text{dim}(\mu ,\nu ).
\end{equation}

Below, we will also be interested in restrictions of the form $S_{n}\times
S_{m-n}\subset S_{m}$, as opposed to $S_{n}\subset S_{m}$ discussed above. The number of times an irrep $(\nu,\nu')$ of $S_{n}\times S_{m-n}$ appears in the restriction of $\mu$ of  $ S_{m}$ is given by the \textit{Littlewood-Richardson coefficients} $g(\mu ;\nu ,\nu
^{\prime })$. These coefficients satisfy the relationship
\begin{equation}\label{relative and LR}
\text{dim}(\mu ,\nu )=\sum_{\nu ^{\prime }\vdash m-n}g(\mu ;\nu ,\nu
^{\prime })\text{dim}_{\nu ^{\prime }}.
\end{equation}

\subsection{The Gelfand-Tsetlin graph}

The analogue of $\mathbb{Y}$ for the unitary groups $U(N)$ goes under the name of \emph{Gelfand-Tsetlin graph} $\mathbb{GT}$. The vertices of $\mathbb{GT}$ correspond to irreps of $U(N)$, and the graph is leveled by the rank of the group $N$. The irreps of $U(N)$ can be labeled by Young diagrams as well. More precisely, at level $N$ we find all the Young diagrams with at most $N$ rows.
 Since there is no bound on the number of columns, there is an infinite number of vertices at each level. The graph grows infinitely upwards as well. Notice that at each level all the diagrams appearing in the lower levels show up. Thus, when referring to a Young diagram as a vertex in $\mathbb{GT}$ one must always specify the level in question, for example $\left(\mu, N\right)\in\mathbb{GT}$.

Now, we introduce the criterion to decide whether two vertices are linked. For the $\mathbb{GT}$ graph this is less straightforward and requires us to introduce some technology. The \textit{signature} of a vertex $\left(\mu, N\right)\in\mathbb{GT}$ is a $N$-tuple of integers, where the first $k$ numbers ($k\leq N$ is the number of rows of $\mu $) are the lengths of the rows of $\mu $ and the rest are 0's; for
example
\begin{equation}
\bigg(\,\begin{Young} &\cr \cr \end{Young},5\,\bigg)\longleftrightarrow
(2,1,0,0,0).
\end{equation}
We say that the signatures of two vertices in $\mathbb{GT}$, $(r_{1},r_{2},\dots ,r_{N})$ and $(s_{1},s_{2},\dots ,s_{N-1})$ at levels $N$ and $N-1$, respectively, \textit{interlace} if and only if
\begin{equation}\label{interlace}
r_{1}\leq s_{1}\leq r_{2}\leq s_{2}\leq \cdots \leq r_{N-1}\leq s_{N-1}\leq
r_{N}.
\end{equation}
Vertices in the Gelfand-Tselin graph are connected if and only if their signatures interlace.
\newline
\newline\newline
\newline
\newline
\newline
\setlength{\unitlength}{2.7cm} {\small
\begin{picture}(3,1)
\put(0,1.1){$N=3$}
\put(1.5,1){$\begin{Young}
&\cr
\cr
\end{Young}$}
\put(0,.5){$N=2$}
\put(1,.5){$\begin{Young}
\cr
\end{Young}$}
\put(1.5,.5){$\begin{Young}
&\cr
\end{Young}$}
\put(2.2,.5){$\begin{Young}
\cr
\cr
\end{Young}$}
\put(2.7,.5){$\begin{Young}
&\cr
\cr
\end{Young}$}
\put(1.5,.95){\line(-1,-1){.3}}
\put(1.6,.95){\line(0,-1){.25}}
\put(1.7,1){\line(1,-1){.35}}
\put(1.8,1.1){\line(3,-1){.8}}
\put(0,-.2){$N=1$}
\put(1,-.2){$\emptyset$}
\put(1.05,0){\line(0,1){.4}}
\put(1.1,0){\line(1,1){.4}}
\put(1.5,-.2){$\begin{Young}
\cr
\end{Young}$}
\put(1.55,0){\line(-1,1){.4}}
\put(1.57,0){\line(0,1){.4}}
\put(1.65,0){\line(1,1){.5}}
\put(1.67,0){\line(2,1){1}}
\put(2,-.2){$\begin{Young}
&\cr
\end{Young}$}
\put(2.1,0){\line(-2,3){.27}}
\put(2.3,0){\line(1,1){.4}}
\put(3,1.3){$\frac{\text{Dim}({\tiny\yng(1,1)}
,2)}{\text{Dim}({\tiny\yng(2,1)}
,3)}$}
\qbezier(2.9,1.2)(1.2384,1.0)
(1.9,0.7722)
\end{picture}} \newline
\newline
\newline

Once more, links form paths in this graph and
as you follow the links all the way to the bottom you move through the restriction chain:
\begin{equation}
U(N)\supset U(N-1)\supset \cdots \supset U(1).  \label{Us}
\end{equation}
As before, the number of downward paths taking from the vertex $(\mu ,N)\in\mathbb{GT}$  to the ground floor equals the dimension of the irrep, $\text{Dim}[\mu ,N]$.
We can define as well the relative dimension
 $\text{Dim}([\mu ,M],[\nu ,N])$, which corresponds to the number of paths (if any) that join
 $[\mu ,M]$ with  $[\nu ,N]$ in the graph\footnote{Hereafter we take $M>N$.}.
The dimensions of irreps of $U(N)$ can be computed using the formula
\begin{equation}\label{UN dim}
\text{Dim}[\nu ,N]=f_{\nu }(N)\frac{\text{dim}_{\nu }}{n!},\quad |\nu |=n,
\end{equation}%
where 
\begin{equation} \label{weight}
f_{\nu }(N)=\prod_{i,j}(N-i+j),
\end{equation}%
known as the \emph{weight}, 
is a product over all the cells in $\nu$ and $\text{dim}_{\nu }$ is given by Eq. \eqref{dim hook}.
Every descending path in $\mathbb{GT}$ can be
represented by a  so-called \textit{Gelfand-Tsetlin pattern}, which are a convenient way of arranging the signatures of the vertices. The interested reader can take a look at the example in appendix \ref{example} to become familiar with these objects.
Finally, since $\mathbb{GT}$ is a leveled graph, probabilities of the form \eqref{proba} can be associated to it. Hence, we introduce
\begin{equation} \label{prob gt}
{}^{\mathbb{GT}}\Lambda _{N}^{M}(\mu ,\nu )= \frac{\text{Dim}[\nu ,N]}{\text{Dim}[\mu ,M]}\text{Dim}([\mu ,M],[\nu ,N])\, ,
\end{equation}%
in analogy with Eq.\,\eqref{prob young}.
Clearly, ${}^{\mathbb{GT}}\Lambda _{N}^{M}(\mu ,\nu )$  satisfies the normalization and compatibility conditions \eqref{nor} and \eqref{compa}.

\subsection{The Young Bouquet and the Borodin-Olshanski identity}

In the previous sections we introduced two leveled graphs, $\mathbb{Y}$ and $\mathbb{GT}$ as well as their associated probability distributions. In spite of the similitudes of these two graphs, they are describing quite different mathematical objects. However, 
one might wonder whether there is any quantitative relationship between them. This question was addressed by Borodin and Olshanski \cite{BO} by comparing the probability distributions \eqref{prob young} and \eqref{prob gt}. 
As a matter of fact, they compared the $\mathbb{GT}$-distribution and a modified version of $\mathbb{Y}$-distribution which we introduce now. A \textit{binomial projective system} is a family of probability distributions
\begin{equation}
{}^{\mathbb{B}}\Lambda _{r}^{r^{\prime }}(m,n)=\Big(1-\frac{r}{r^{\prime }}%
\Big)^{m-n}\Big(\frac{r}{r^{\prime }}\Big)^{n}\frac{m!}{(m-n)!n!},
\label{binomial}
\end{equation}%
where $r,r^{\prime }\in\mathbb{R}^+$, $r^{\prime }>r$  and $n,m$ are non-negative
integers. One can readily check that the
compatibility condition
\begin{equation}
\sum_{l} {}^{\mathbb{B}}\Lambda _{r^{\prime \prime }}^{r^{\prime }}(m,l ){}^{\mathbb{B}%
}\Lambda _{r}^{r^{\prime \prime }}(l,n)={}^{\mathbb{B}}\Lambda _{r}^{r^{\prime }}(m,n)\,,
\label{YBcom}
\end{equation}%
for any intermediate level, i.e. $r<r''<r'$,
is satisfied. By combining \eqref{binomial} with \eqref{prob young}, Borodin and Olshanski define the \textit{Young
Bouquet} ($\mathbb{YB}$) whose associated distribution reads
\begin{equation}
{}^{\mathbb{YB}}\Lambda _{r}^{r^{\prime }}(\mu ,\nu )=\Big(1-\frac{r}{%
r^{\prime }}\Big)^{m-n}\Big(\frac{r}{r^{\prime }}\Big)^{n}\frac{m!}{(m-n)!n!}%
\frac{\text{dim}_{\nu }}{\text{dim}_{\mu }} \text{ dim}(\mu ,\nu ),\label{YB dist}
\end{equation}%
where in the above $|\mu |=m$ and $|\nu |=n$, and $m\geq n$.

It is the Young Bouquet, which is found to have a deep
connection with the $\mathbb{GT}$ graph. The identity found by Borodin and Olshanski reads \cite{BO}
\begin{equation}
\lim_{\frac{N}{M}\rightarrow \frac{r}{r^{\prime }}}{}^{\mathbb{GT}}\Lambda
_{N}^{M}([\mu ,M],[\nu ,N])={}^{\mathbb{YB}}\Lambda _{r}^{r^{\prime }}(\mu
,\nu ),  \label{sim}
\end{equation}%
where $N,M\rightarrow \infty$ and $N/M$ fixed. Formula (\ref{sim})
is a deep mathematical identity which depends only on how the branching
graphs and their boundaries ($M,N\rightarrow \infty$) are constructed which,
in the end, it depends on how irreps of the groups are subduced  \cite{BO,BO2012}. Henceforth,
we refer to Eq.\,\eqref{sim} as the BO identity or $\mathbb{YB}/\mathbb{GT}$ duality. One of the results of the present
work is to provide a succinct demonstration of Eq.\,\eqref{sim}.

\section{Schur transitions and the $\mathbb{YB}/\mathbb{GT}$ duality}

\label{sec: multi-graviton transitions}

\begin{figure}[h!]
\centering
\includegraphics[trim=0cm 1.5cm 0cm 1.6cm, clip=true,scale=0.6]{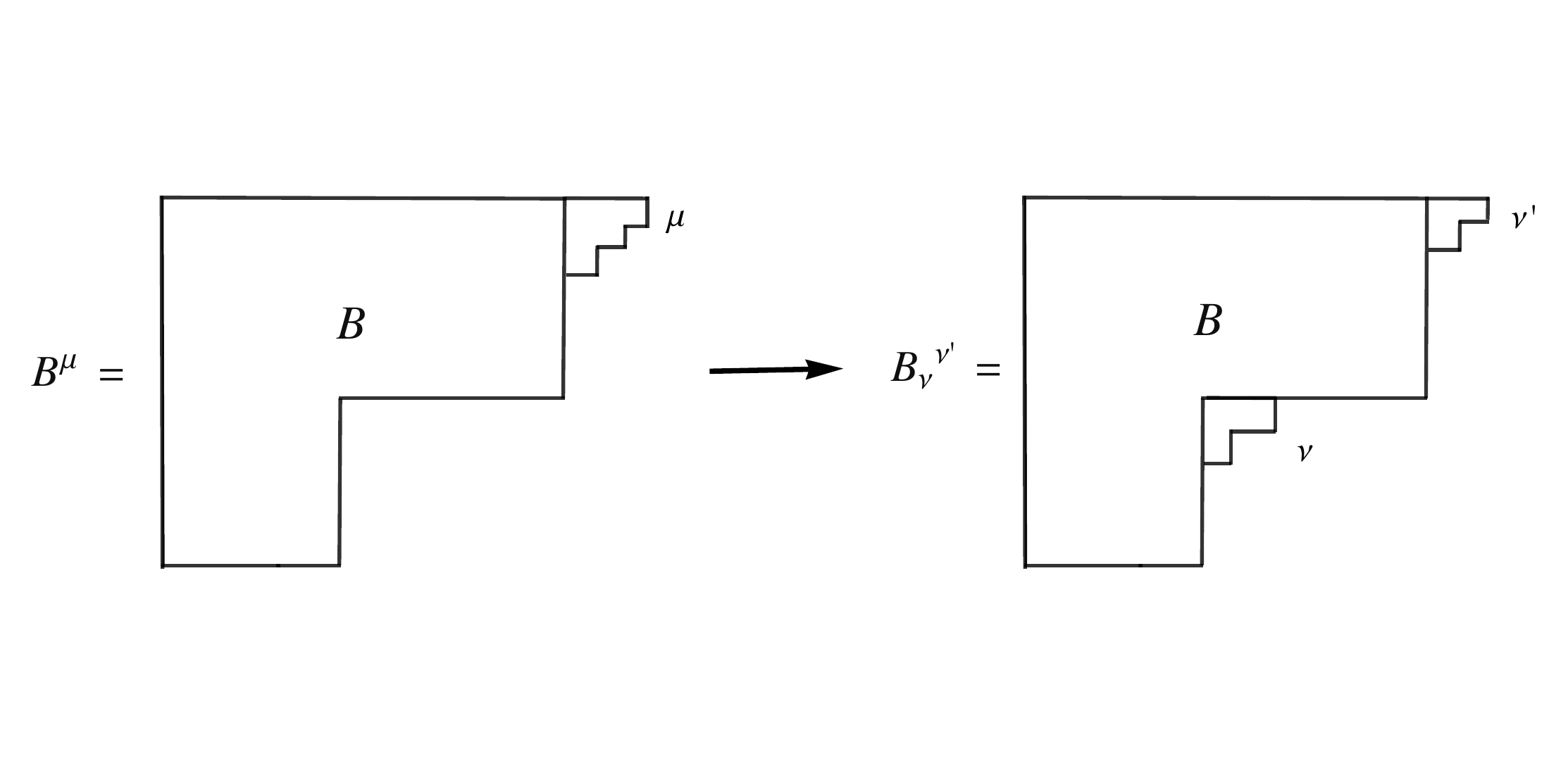}
        \caption{The Young diagram $\mu$ is placed in the top left corner, then its boxes are redistributed forming the diagrams $\nu$ and $\nu'$.}\label{Transition}
\end{figure}

\begin{figure}[h!]
\centering
\includegraphics[trim=2cm 2cm 4cm 1cm, clip=true,scale=0.43]{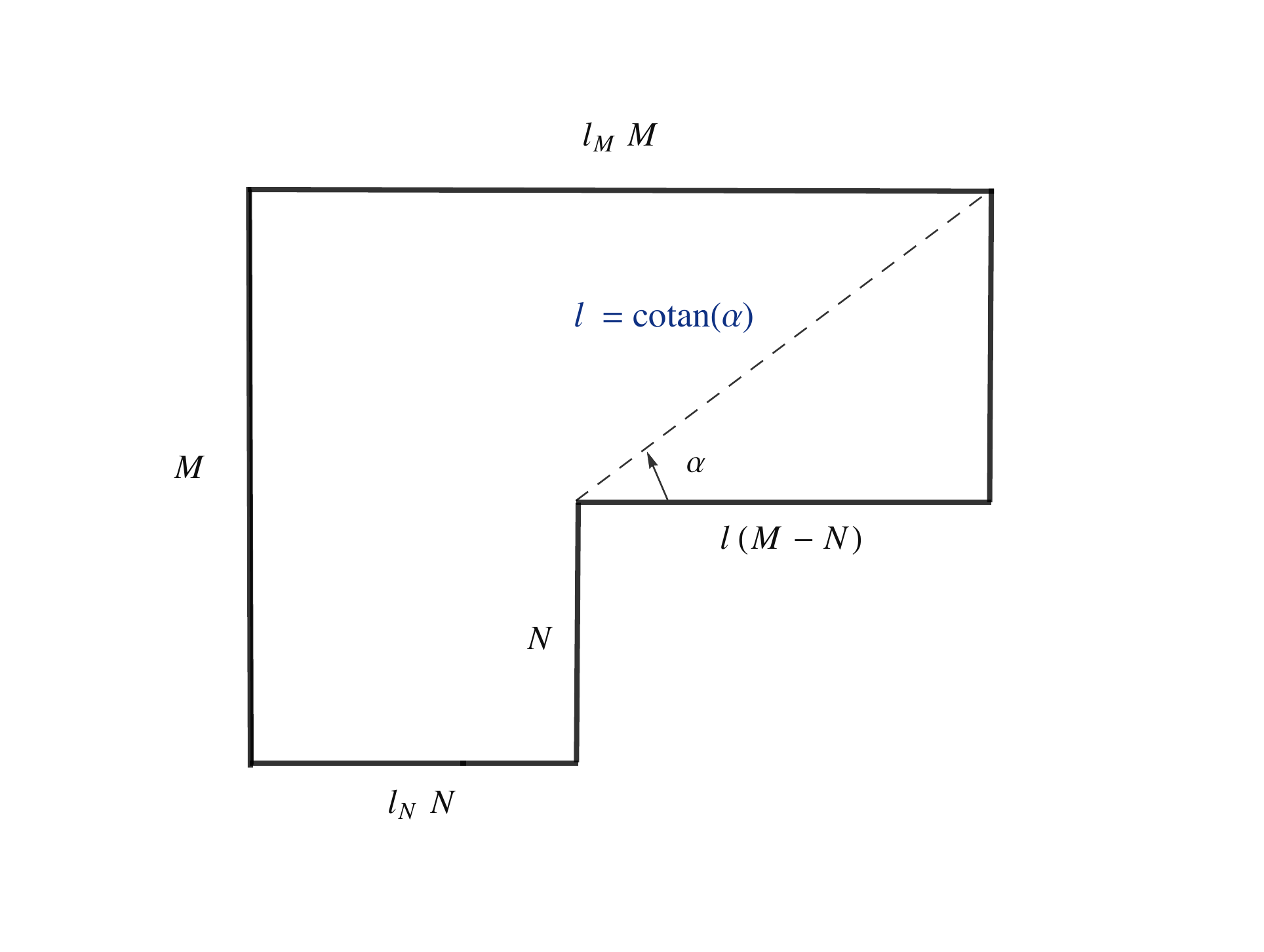}
        \caption{The hook-shaped or two-ring geometry where all the side lengths have been explicitly written. We will consider large $l$, which accounts for thin and long hook shaped backgrounds.     }\label{cot}
\end{figure}

In the present section, inspired by computations of transition probabilities in $U(M)$ ${\cal N}=4$ SYM we calculate probabilites of processes such as the one depicted in  Fig.\,\ref{Transition} where the background is given by Fig.\,\ref{cot}. In previous work \cite{Diaz:2015tda}, it was shown that for large gauge groups these probabilities match the Young bouquet's distribution ${}^{\mathbb{YB}}\Lambda _{r}^{r^{\prime }}(\mu
,\nu )$.
 Below, we revisit these processes at finite $N$ and $M$. Our findings lead us to conjecture a compelling expression for the relative dimensions of unitary groups. The proof of this conjecture will be provided in the next section. Moreover,
we argue that using this expression, the $\mathbb{YB}/\mathbb{GT}$ duality Eq.\,\eqref{sim} can be easily deduced.

We will refer to processes such as the one portrayed in Fig.\,\ref{Transition} as \emph{Schur transitions} or \emph{multigraviton transitions} .
The latter terminology has its roots in the AdS/CFT correspondence \cite{Maldacena:1997re}, more precisely, in the work of Lin, Lunin and Maldacena (LLM) \cite{Lin:2004nb}. The LLM prescription allows one to construct a metric with an AAdS factor out of a large Young diagram such as $B$. In particular, under this procedure $B$ would give rise to a domain wall geometry in the bulk (see \cite{Diaz:2015tda}  for more details). Moreover, under suitable circumstances, the small diagrams $\mu$ , $\nu$ and $\nu'$ can be thought of as multigraviton excitations on the background generated by $B$ \cite{Koch:2008ah}. Although we consider these transitions in a different context, we will still call them multigraviton transitions.
Hereafter, we  will refer to the upper-rightmost corner of this \textit{hook-shaped diagram} as the \textit{$M$-corner}, and to the inward pointing corner as the \textit{$N$-corner}.

\begin{figure}
        \centering

        \begin{subfigure}[h]{0.5\textwidth}
\hspace{19mm}\includegraphics[scale=0.25]{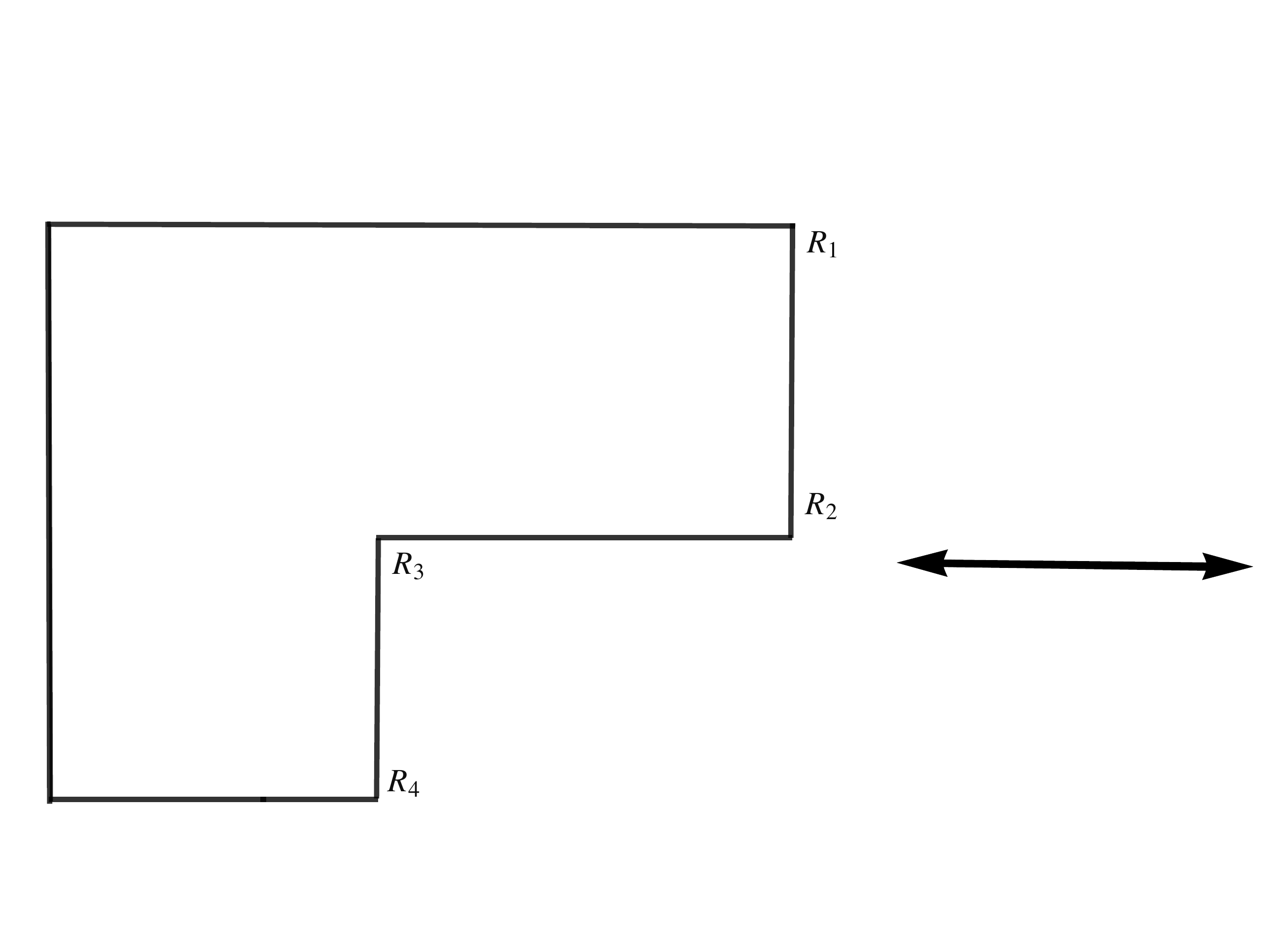}
                      \end{subfigure}%
        ~
    \begin{subfigure}[h]{0.5\textwidth}
\hspace{2mm}\includegraphics[scale=0.26]{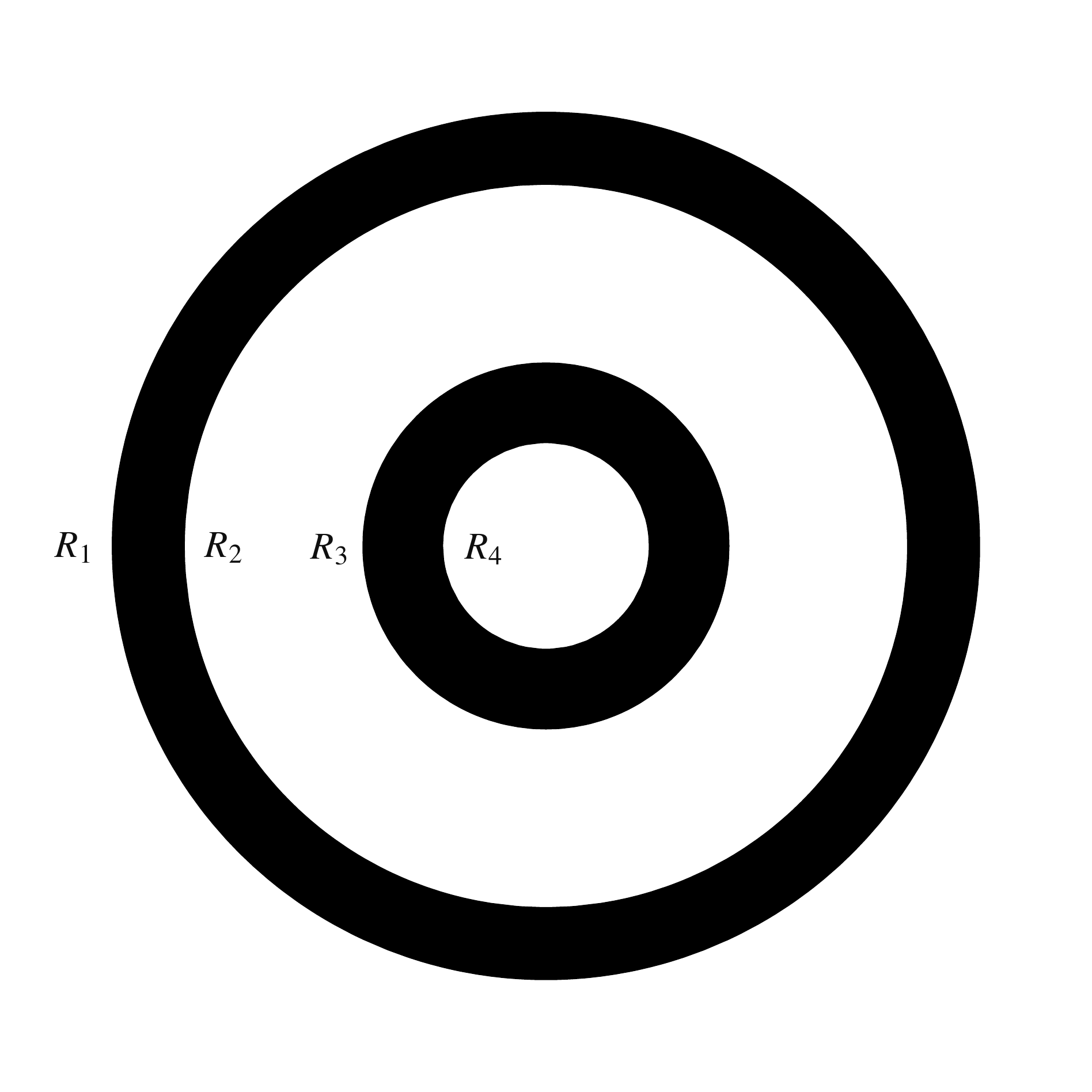}
        \end{subfigure}
        \caption{One-to-one relation between Young diagrams and the bubbling plane.}\label{Young bubbles}
\end{figure}

In practice, the state corresponding to Fig.\,\ref{cot} is produced by acting on the vacuum with a Schur polynomial
\begin{equation}
\chi _{B}(Z)|0\rangle =|B\rangle \,,
\end{equation}
constructed out of the scalar field $Z$ of the ${\cal N}=4$ gauge theory. By attaching a Young diagram $\mu$ or $\nu$ either to the $M$-corner or to the $N$-corner, we create multigraviton states in the outer edges of the rings in Fig. \ref{Young bubbles}.
 The process whose amplitude we consider is described in Fig. \ref{Transition}, where the number of boxes is conserved, namely
\begin{equation}
|\mu |=m,\quad |\nu |=n,\quad |\nu ^{\prime }|=m-n\,.
\end{equation}%
We denote the probability of this transition by
\begin{equation}
P_{\nu }^{\mu \nu
^{\prime }}={\cal P}(B^{\mu }\rightarrow B_{\nu }^{\nu ^{\prime }})\,.\label{MG}
\end{equation}


Moreover, we want the interaction between the multigravitons and the background to be purely gravitational. Therefore, we must consider excitations with vanishing angular momentum in the $Z$ direction, so the multigravitons must be constructed using a field in the theory different from $Z$. Let us use $Y$ for that purpose. Multigraviton states are also half-BPS and as such they are given by Schur polynomials $\chi _{\mu }(Y)$ and $\chi _{\nu }(Y)$, where $\mu $ and $\nu $ are Young diagrams with $m$
and $n$ boxes, respectively. The product of background and excitation can be
written in terms of restricted Schur polynomials as \cite{Bhattacharyya:2008xy}
\begin{equation}
\chi _{B}(Z)\chi _{\mu }(Y)=H_{B}H_{\mu }\sum_{B_{\nu
}^{\nu ^{\prime }},i}\frac{1}{H_{B_{\nu }^{\nu ^{\prime }}}}\chi
_{B_{\nu }^{\nu ^{\prime }},(B,\mu )^{i}}(Z,Y),  \label{partonsproduct}
\end{equation}
where the $B_{\nu }^{\nu ^{\prime }}$ are diagrams that can be formed from
the product $B\times \mu $, and $i$ runs over the multiplicities
given by the Littlewood-Richardson coefficients $g(B_{\nu }^{\nu ^{\prime}};B,\mu )$. \footnote{See appendix \ref{ResSchur} for a discussion}


In terms of correlators of Schur polynomials, the multigraviton transition probability $P_{\nu }^{\mu \nu^{\prime }}$ reads
\begin{equation}
P_{\nu }^{\mu \nu
^{\prime }}=\frac{|\langle \chi
_{B}^{\dagger }(Z)\chi _{\mu }^{\dagger }(Y)\chi _{B_{\nu }^{\nu ^{\prime
}},(B,\mu )}(Z,Y)\rangle |^{2}}{\Vert \chi _{B}(Z)\chi _{\mu }(Y)\Vert
^{2}\Vert \chi _{B_{\nu }^{\nu ^{\prime }},(B,\mu )}(Z,Y)\Vert ^{2}}.\label{prob}
\end{equation}
Actually, we know that the Young bouquet distribution emerges from the sum over intermediate states of the above expression, i.e. $P_{\nu }^{\mu}\equiv{\cal P}(B^{\mu }\rightarrow B_{\nu })$. Explicitly, this probability is given by
\begin{equation}
P_{\nu }^{\mu}=\sum_{\nu ^{\prime }}\frac{|\langle
\chi _{B}^{\dagger }(Z)\chi _{\mu }^{\dagger }(Y)\chi _{B_{\nu }^{\nu
^{\prime }},(B,\mu )}(Z,Y)\rangle |^{2}}{\Vert \chi _{B}(Z)\chi _{\mu
}(Y)\Vert ^{2}\Vert \chi _{B_{\nu }^{\nu ^{\prime }},(B,\mu )}(Z,Y)\Vert ^{2}%
}.  \label{probabilities}
\end{equation}
This is the quantity that we compute in the following.

Recall that the two-point function of restricted Schur operators is given by \cite{Bhattacharyya:2008rb}
\begin{equation}
\langle \chi _{R,(r,s)}^{\dagger }(Z,Y)\chi _{T,(t,u)}(Z,Y)\rangle =\delta
_{RT}\delta _{rt}\delta _{su}\frac{H_{R}}{H_{r}H_{s}}f_{R}\, .  \label{restricted}
\end{equation}%
Using Eqs.\eqref{partonsproduct} and \eqref{restricted} we can compute all the quantities appearing in Eq.\,(\ref%
{probabilities})
\begin{eqnarray}\label{parts}
\Vert \chi _{B}(Z)\chi _{\mu }(Y)\Vert ^{2} &=&\langle \chi _{B}^{\dagger
}(Z)\chi _{\mu }^{\dagger }(Y)\chi _{B}(Z)\chi _{\mu }(Y)\rangle
=f_{B}f_{\mu },  \notag \\
\Vert \chi _{B_{\nu }^{\nu ^{\prime }},(B,\mu )}(Z,Y)\Vert ^{2} &=&\langle
\chi _{B_{\nu }^{\nu ^{\prime }},(B,\mu )}^{\dagger }(Z,Y)\chi _{B_{\nu
}^{\nu ^{\prime }},(B,\mu )}(Z,Y)\rangle =\frac{H_{B_{\nu }^{\nu
^{\prime }}}}{H_{B}H_{\mu }}f_{B_{\nu }^{\nu ^{\prime
}}},  \notag \\
|\langle \chi _{B}^{\dagger }(Z)\chi _{\mu }^{\dagger }(Y)\chi _{B_{\nu
}^{\nu ^{\prime }},(B,\mu )}(Z,Y)\rangle |^{2} &=&f_{B_{\nu }^{\nu ^{\prime
}}}^{2}g(\mu ;\nu ,\nu ^{\prime }).
\end{eqnarray}%
Above $f_{B}$, $f_{\mu }$, and $f_{B_{\nu }^{\nu ^{\prime}}}$ stand for the weights (Eq.\,\eqref{weight}) of the Young diagrams $B$, $\mu$, and $B_{\nu }^{\nu ^{\prime}}$ respectively. 

\begin{figure}[h]
    \centering
\includegraphics[trim =0cm 4cm 2cm 2cm, clip, scale=0.42]{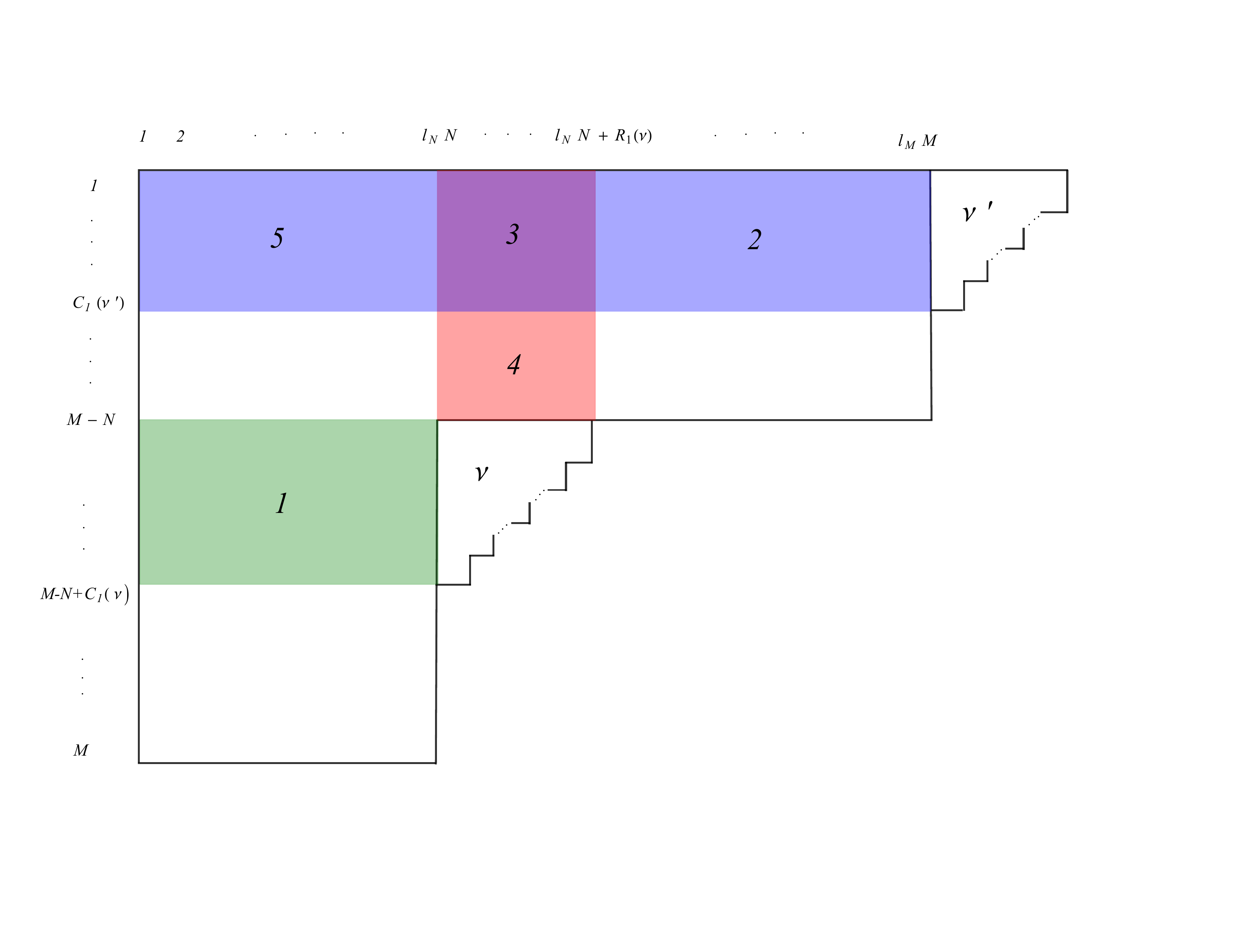}
    \caption{The colored regions in $B_{\nu }^{\nu ^{\prime
}}$ contributing to Eq.\,\eqref{exact Pmn}. Let $R_I(\nu)$ stand for the length of the $I$th row in $\nu$ and $C_J(\nu)$ for the height of the $J$th column of $\nu$, with similar definitions for $\nu'$.
Region $(1)$ comprises the area between rows $M-N$ and $M-N+C_1(\nu)$, and columns $1$ to $l_N N$; region (2), rows $1$ to $C_1(\nu')$ and columns $l_N N+R_1(\nu)+1$ and $l_MM$; region (3) same rows as (2) but columns $l_N+1$ to $l_NN+R_1(\nu)$; region (4) rows $C_1(\nu')$ to $M-N$ and same columns as (3); region (5) has the same rows as (2) and columns from $1$ to $l_N N$. }
    \label{fig:regions1}
\end{figure}

It can be shown that for $B$ as in Fig. \ref{Young bubbles} we have $g(B_{\nu }^{\nu ^{\prime
}};B,\mu )=g(\mu ;\nu ,\nu ^{\prime })$. Using this fact together with Eqs.\,\eqref{parts}, the expression (\ref%
{probabilities}) reads
\begin{equation}
P_{\nu }^{\mu}=\sum_{\nu ^{\prime }\vdash m-n}g(\mu ;\nu
,\nu ^{\prime })\frac{f_{B_{\nu }^{\nu ^{\prime }}}}{f_{B}f_{\mu }}\frac{%
H_{B}H_{\mu }}{H_{B_{\nu }^{\nu ^{\prime }}}\label{exact Pmn}
}.
\end{equation}%
Observe that up until now we have made no approximations, thus, the result in Eq. \eqref{exact Pmn} is exact. 
As shown in \cite{Diaz:2015tda}, in the large $N$, $M$  and $l$ limits \eqref{probabilities} reproduces the probability distribution of the Young bouquet ($\mathbb{YB}$). 
In the following calculation we explore what happens if we drop the large $M$ and $N$ assumption. For this computation, it will prove convenient for the reader to have Fig.\,\ref{fig:regions1} in sight. 
First of all, for the hook-lengths we have
\begin{align}
\frac{H_{B}H_{\mu }}{H_{B^{\nu '}_{\nu }}}=&
\frac{H_{\mu }}{H_{\nu }H_{\nu' }}\prod_{\kappa=1}^5 \frac{H^{(\kappa)}_{B}}{H^{(\kappa)}_{B^{\nu '}_{\nu }}}\nonumber\\
=&\frac{m!}{(m-n)!n!}\frac{\text{dim}_{\nu ^{\prime }}\text{dim}_{\nu }}{\text{%
dim}_{\mu }}
\prod_{\kappa=1}^5 \frac{H^{(\kappa)}_{B}}{H^{(\kappa)}_{B^{\nu '}_{\nu }}}\label{eq:just hooks 2}\,,
\end{align}
where $H^{(\kappa)}_{B}$ stands for the hook-length contributions coming from the colored regions in Fig.\,\ref{fig:regions1} and $\kappa=1,\dots,5$. Meanwhile, the contribution of the weights is given by 
\be
\frac{f_{B_{\nu }^{\nu ^{\prime }}}}{f_{B}f_{\mu }}=\frac{f_{\nu}[N(l_N+1)]f_{\nu'}[M(l_M+1)]}{f_{\mu}[M]}.
\ee

\noindent In the large $l$ limit, the only non-trivial contributions to Eq. \eqref{eq:just hooks 2} come from regions $\kappa=1,2$. In particular, region $\kappa=1$ contributes
\begin{align}
\frac{H^{(1)}_{B}}{H^{(1)}_{B^{\nu '}_{\nu }}}= &\prod_{I=1}^{C_1(\nu)}\prod_{j=1}^{R_I(\nu)}\frac{N-I-j+R_I(\nu)+1}{N(l_N+1)-I-j+R_I(\nu)+1}\,,
\end{align}
where $R_I(\nu)$ is the length of the $I$th row in $\nu$ and $C_J(\nu)$  the height of the $J$th column of $\nu$ (see Fig.\,\ref{fig:regions1}). From this, we find
\be
\frac{H^{(1)}_{B}}{H^{(1)}_{B^{\nu '}_{\nu }}}= \frac{f_\nu[N]}{f_\nu[N(l_N+1)]}\,.
\ee

Meanwhile, region $\kappa=2$ yields
\be
\lim_{l\to\infty} \frac{H^{(2)}_{B}}{H^{(2)}_{B^{\nu '}_{\nu }}}\to f_{\nu'}[M-N]\,[l(M-N)]^{n-m}.
\ee
In turn, for the weight contributions we find  
\be
\lim_{l\to\infty}f_{\nu'}[M(l_M+1)]\to [l(M-N)]^{m-n}\,.
\ee
Hence, it follows that 
\be
\lim_{l\to\infty}\frac{f_{B_{\nu }^{\nu ^{\prime }}}}{f_{B}f_{\mu }}\frac{H_{B}H_{\mu }}{H_{B^{\nu '}_{\nu }}}=\frac{m!}{(m-n)!n!}\frac{\text{dim}_{\nu ^{\prime }}\text{dim}_{\nu }}{\text{%
dim}_{\mu }}\frac{f_{\nu}[N]f_{\nu'}[M-N]}{f_{\mu}[M]}\,.
\ee
Therefore, Eq.\,\eqref{exact Pmn} reads
\be
\hat P_\nu^{\mu}\equiv \lim_{l\to\infty}P_\nu^{\mu}= \frac{m!}{n!(m-n)!}\left(\frac{f_\nu[N]\,\text{dim}_{\nu } }{f_\mu[M]\,\text{dim}_{\mu }}\right) \sum_{\nu ^{\prime }\vdash m-n} g(\mu ;\nu,\nu ^{\prime }) \text{dim}_{\nu' }f_{\nu'}[M-N]. \label{eq:finite}
\ee
This formula will be central in the following discussion. 

Let us consider two interesting limits of the above expression. First, taking $N,M\to\infty$ with $M/N$ fixed, Eq.\,\eqref{eq:finite} goes like
\begin{align}
\hat P_\nu^{\mu}
\sim&\binom{m}{n}\Big(1-\frac{N%
}{M}\Big)^{m-n}\Big(\frac{N}{M}\Big)^{n}\frac{\text{dim}%
_{\nu }}{\text{dim}_{\mu }} \text{ dim}(\mu ,\nu )\, .
\end{align}
Thus, upon the identification $N/M\leftrightarrow r'/r$ , we recover the Young bouquet distribution Eq.\,\eqref{YB dist}. Moreover, since we are in the large $M,N$  regime, the $\mathbb{YB}/\mathbb{GT}$ duality Eq.\,\eqref{sim} implies  
\be 
\hat P_\nu^{\mu}={}^{\mathbb{GT}}\Lambda_{N}^{M}([\mu ,M],[\nu ,N])\,.
\ee
Alternatively, we could keep $M$ and $N$ fixed and take $\mu=\nu$. In this scenario we have 
\be
\hat P_{\mu}^{\,\mu}=\frac{f_\mu\left[N\right]}{f_\mu\left[M\right]}\,.\label{charges}
\ee
The key point is, that the above expression also corresponds to the $\mathbb{GT}$ distributions. Since these are rather different regimes we are led to conjecture that 
\be\label{conjecture}
\hat P_{\nu}^{\,\mu}={}^{\mathbb{GT}}\Lambda
_{N}^{M}([\mu ,M],[\nu ,N])\,,
\ee
holds in general.
Using Eqs. \eqref{UN dim}, \eqref{prob gt} and \eqref{eq:finite}, this claim can be expressed succinctly as a formula for the unitary group's relative dimensions
\begin{eqnarray}
\text{Dim}[\mu,M;\nu,N]
=\sum_{\nu ^{\prime }\vdash m-n} g(\mu ;\nu,\nu ^{\prime })\text{Dim}[\nu^{\prime}, M-N]\,. \label{relativedimension}
\end{eqnarray}
Examples verifying the correctness of this expression are provided in appendix \ref{examples}. In the following section we give a proof of Eq.\,\eqref{relativedimension} for the relative dimension. 
As we shall see, given formula \eqref{relativedimension} the $\mathbb{YB}/\mathbb{GT}$ duality  Eq. \eqref{sim} follows as a simple corollary.

It is worth to mention that the branching graph rules, and so the above formulas, are identical for $SU(N)$ groups. The interlacing condition is the same. The difference between the $SU(N)$  and $U(N)$ branching graphs is that at level $N$, only Young diagrams with at most $N-1$ rows are allowed for $SU(N)$ as opposed to a maximum of $N$ rows that allows $U(N)$ irreps. In other words, irreps of $SU(N)$ are characterized by signatures with $N-1$ integers. Clearly the $SU(N)$ branching graph is a subgraph of the $U(N)$ graph.  


\section{A proof of the relative dimension formula}\label{proof}

Recall that the $\mathbb{GT}$ graph organizes the irreps of unitary groups such that the number of paths descending from a vertex to the bottom matches the dimension of the irrep. To compute this number, we might use the formula \eqref{UN dim}. Alternatively, we could count the number of triangular patterns formed by interlacing signatures like those in Eq.\,\eqref{GT patterns}.
As we have seen, the use of leveled graphs allows us to introduce the concept of \emph{relative dimension}; which counts downward paths connecting two irreps e.g. $[\mu,M]$ and $[\nu,N]$. There is no formula analogous to \eqref{UN dim} to compute the relative dimension immediately. Rather, we are compelled to perform a direct counting of GT patterns; this time, truncated in such a way that the last line corresponds to the signature of $[\nu,N]$. We refer to these patterns as \emph{partial} or \emph{trapezoidal} GT patterns. In general, carrying out this enumeration is quite cumbersome. At the end of the previous section, we claimed that 
the relative dimension can be computed using formula \eqref{relativedimension} instead. A number of examples of how this formula works can be found in appendix \ref{examples}. In this section, we provide a proof for our claim.

Our demonstration follows easily using some basic Schur function's technology, appendix \ref{schur f} contains a summary of the relevant tools. The key point is that
Schur functions evaluated at $x_i=1$ yield the dimension of the irreps of unitary groups \cite{Mac}, that is
\be
S_{\mu}(\underbrace{1,1,\dots,1}_M)=\text{Dim}[\mu,M].
\ee
In turn,  the relative dimension can be written as
\be 
S_{\mu/\nu}(\underbrace{1,1,\dots,1}_{M-N})=\text{Dim}[\mu,M;\nu, N],
\ee
where $S_{\mu/\nu}$ are skew Schur functions. The product of Schur functions can be written in terms of Littlewood-Richardson coefficients as 
\be \label{prod}
S_\nu S_{\nu'} =\sum_{\mu}g(\mu;\nu,\nu')S_\mu\,.
\ee
From this product rule and Eq.\,\eqref{skew} we find that
\be
S_{\mu/\nu}=\sum_{\nu'}g(\mu;\nu,\nu')S_{\nu'}.
\ee
Writing out the variables
\be
S_{\mu/\nu}(x_1,\dots, x_{M-N})=\sum_{\nu'}g(\mu;\nu,\nu')S_{\nu'}(x_1,\dots, x_{M-N})\, ,
\ee
and then setting them to 1 yields
\be
\text{Dim}[\mu,M;\nu, N]=\sum_{\nu'}g(\mu;\nu,\nu')\text{Dim}[\nu', M- N]\,,
\ee
which is identical to (\ref{conjecture}),
thus, proving our claim.
Observe that there is a appealing parallelism between relative dimensions of the both graphs mediated by the Littlewood-Richardson coefficients, see Eq.\,\eqref{relative and LR}.

As an application of this result, we provide a straightforward proof of the $\mathbb{YB}/\mathbb{GT}$ duality \eqref{sim}. It goes as follows, since 
\begin{align}
{}^{\mathbb{GT}}\Lambda^M_N(\mu,\nu)&=\left(\frac{\text{Dim}[\nu,N]}{\text{Dim}[\mu,M]}\right)\text{Dim}[\mu,M;\nu,N]\nn\\
&= \left(\frac{\text{Dim}[\nu,N]}{\text{Dim}[\mu,M]}\right) \sum_{\nu'}g(\mu;\nu,\nu')\text{Dim}[\nu', M-N]\,,
\end{align}
then, Eq.\,\eqref{UN dim} implies
\be\label{exact}
^{\mathbb{GT}}\Lambda^M_N(\mu,\nu)=\frac{m!}{n!(m-n)!}\frac{\text{dim}_{\nu}}{\text{dim}_{\mu}}\frac{f_\nu(N)}{f_\mu(M)}\sum_{\nu'}g(\mu;\nu,\nu')f_{\nu'}(M-N)  \text{dim}_{\nu'}\,.
\ee
Finally, considering the limit $M,N\rightarrow \infty$ we find 
\be
^{\mathbb{GT}}\Lambda^M_N(\mu,\nu)=\frac{m!}{n!(m-n)!}\bigg(1-\frac{N}{M}\bigg)^{m-n}\bigg(\frac{N}{M}\bigg)^{n} \frac{\text{dim}_{\nu}}{\text{dim}_{\mu}}\,\text{dim}(\mu,\nu)\, ,
\ee
where we made use of Eq.\,\eqref{relative and LR}. Clearly, the above expression matches the Young bouquet's distribution \eqref{YB dist}
upon making the identification $N/M\rightarrow r/r'$; this proves Eq.\,\eqref{sim}. Notice that one of the key steps in this computation is that the leading term of the factor $f_{\nu'}(M-N)$ in Eq.\,\eqref{exact} is independent of the shape of $\nu'$. It would be interesting to explore what happens as one considers subleading contributions to this formula, we leave that question for future work. 

Once more, we would like to remind the reader that all the above formulas calculated for the $U(N)$ branching graph also hold for the $SU(N)$ branching graph. The latter being a subgraph of the $U(N)$ graph. The branching (or interlacing) condition is exactly the same. The only difference is that irreps of $SU(N)$ are labeled by Young diagrams with at most $N-1$ instead of $N$ rows.


\section{Towards a quantum relative dimension?}\label{QRD}

In the context of rational conformal field theories (RCFTs)\footnote{ See \cite{Moore:1989vd} for an excellent review on the subject.} there is a natural generalization of the notion of the dimension 
of an irreducible representation. RCFTs are characterized by the fact that the operator content of the theory can be organized into a finite number 
of families. This class contains very interesting systems such as minimal models and Wess-Zumino-Witten (WZW) theories. In the following, we shall concern ourselves mainly with the latter type. These models are endowed, alongside the Virasoro symmetry algebra, with currents that conform to an affine Lie algebra structure $\mathfrak{g}_k$
\be
\left[j_m^a, j_n^b \right]=i\sum_c f_{abc} j_{m+n}^c+k\, m\,\delta^{ab}\,\delta_{m+n}\, ,
\ee
where $f_{abc}$ are the structure constants of the Lie algebra $\mathfrak{g}$, and $k$ is the so-called \emph{level} of the model. Although these theories generally have an infinite number of Virasoro families, the families can be rearranged into a finite number of affine Lie algebra representations. To each such family, it is possible to associate a character defined as
\be
\chi_\mu(\tau)=\Tr_{\hspace{-1mm}\mu}\,e^{2\pi i \tau\left(L_0-c/24\right)}\,,
\ee
where $\mu$ is a label for the affine primary.

 The \emph{quantum dimension} of a family is defined as
 \be
 \hat {d_\mu}(\mathfrak{g}_k)=\lim_{\tau\to i 0^+}\frac{\chi_\mu(\tau)}{\chi_0(\tau)}\,,
 \ee
and it estimates the \emph{size} with respect to the vacuum of the Hilbert space associated with an affine family. A central fact in the study of RCFTs is that under modular transformations the characters transform linearly as
\be\label{mod S}
\chi_\mu\left(-1/\tau\right)=\sum_\lambda {\cal S}_{\mu\lambda}\,\chi_\lambda(\tau)\,.
\ee
The matrix $ {\cal S}$ is commonly known as the \emph{modular S-matrix}. Many important quantities and structures of the RCFT are encoded in $ {\cal S}$; the quantum dimension is no exception, indeed we have
\be
\hat {d_\mu}(\mathfrak{g}_k)= \frac{{\cal S}_{\mu\, 0}}{{\cal S}_{ 0 0}}\,.
\ee
Besides their own intrinsic interests RCFTs, it is worth studying them due to 
their relationship to Chern-Simons theory \cite{Witten:1988hf} and topological phases of matter such as fractional quantum Hall  fluids \cite{Wen:1990se}. Moreover, both the quantum dimension and the modular S-matrix have appeared in a number of recent works. For example, in \cite{He:2014mwa, Caputa:2015tua} it was shown that after inserting any operator in the family of $\mu$, the R\'enyi entropy of the system jumps  by 
\be
\Delta S^{(n)}=\log (\hat {d_\mu})\,.
\ee
More recently, it was discovered that a particular combination of these quantities, called the \emph{anyon monodromy}
\be
{\cal C}_{\mu\nu}=\frac{1}{\hat {d_\mu}\hat {d_\nu}}\frac{{\cal S}^*_{\mu\nu}}{{\cal S}_{00}}
\ee
serves as a diagnostic of quantum chaos \cite{Caputa:2016tgt, Gu:2016hoy}.

For concreteness and in order to make contact with the previous section let us consider level-$k$ WZW models where the underlying Lie algebra is $SU(N)$. The affine primaries in these models, are in one to one correspondence with Young diagrams with less than $N$ rows and at most $k$ columns. Before proceeding, we introduce the so-called \emph{$q$-numbers} (which can be defined for any affine Lie algebra)
\be
\left[x\right]=\frac{q^{x/2}-q^{-x/2}}{q^{1/2}-q^{-1/2}}\hspace{9mm} q=\exp\left(-\frac{2\pi i}{k+C_{\mathfrak{g}}}\right)\, ,
\ee
where $C_{\mathfrak{g}}$ is the dual Coxeter number of the Lie algebra. It is crucial to notice that 
\be\label{classical}
  \lim_{k\to\infty}[x]=x \,,
\ee
this limit is known as the classical limit of the WZW model.
 For the case at hand, namely $SU(N)_k$ theories, $C_{\mathfrak{g}}=N$.
 The quantum dimensions in $SU(N)_k$ models can be written succinctly
 in terms of the Schur function 
\be\label{Schur dim}
 S_\mu\left(x_1,x_2,\dots,\, x_N\right)=\frac{\det\left[  x_j^{l(\mu)_i+N-i}\right]}{\det\left[  x_j^{N-i}  \right]}\hspace{6mm} j=1,\dots,\, N\,,
\ee
as \cite{Dong:2008ft}
 \be\label{dim Schur}
\hat d_{\mu}(N)=q^{-N(N-1)\kappa_\mu/2}\, S_\mu\left(q^{N-1},q^{N-2},\dots, \, 1\right)\,.
\ee
In equation \eqref{dim Schur} we introduced the quantity 
\be
\kappa_\mu=\frac{1}{N}\sum_{j=1}^{N-1} \mu_i\,,
\ee
where the $ \mu_i$ are the Dynkin labels of $\mu$. Moreover, in equation \eqref{Schur dim} $l_i(\mu)$ stands for the number of boxes in the $i$th row of $\mu$'s Young diagram. 
Finally, the Schur function \eqref{Schur dim} can be rexpressed conveniently in terms of the quantities  
\be \label{generating}
E_{0<m<N}=q^{\frac{(N-1) m}{2}}\prod_{l=1}^m \frac{\left[ N+1-l  \right]}{\left[ l \right]}\,,\hspace{5mm} E_0=1\,,
\ee
 as 
\be\label{Schur 2}
  S_\mu\left(q^{N-1},q^{N-2},\dots, \, 1\right)=\det(E_{\mu^{T}_i+j-i})\,,
\ee

 In \cite{Dong:2008ft} the quantum dimensions of the fundamental, symmetric and antisymmetric representations are computed explicitly and they read
\begin{align}\label{dimensions}
&\begin{Young}\cr
\end{Young}
\hspace{12mm}  \hat d_\mu(N)=[N] \nn\\
&\begin{Young}\cr\cr
\end{Young}
\hspace{12mm}  \hat d_\mu(N)=\frac{[N][N+1]}{[2]}\nn\\
&\begin{Young}&\cr
\end{Young}
\hspace{7mm}  \hat d_\mu(N)=\frac{[N][N-1]}{[2]}\,.
\end{align}
Now, we present another example for future use and in order to familiarize the reader with the computation. Let $\mu$ be the adjoint representation, which is labeled by the Young diagram
\be 
\begin{Young}&\cr
\cr \end{Young}\hspace{9mm}
\ee
for which we have $\mu^{T}=(2,1,0,\dots,0)$. Thus, the matrix in Eq.\,\eqref{Schur 2} reads
\be
(E_{\mu^{T}_i+j-i})=\left( \begin{array}{ccccc}
E_2 & E_3 & \dots & \dots &  E_{N+1}\\
E_0 & E_1 & E_2& \dots&  E_{N-1}\\
0 & 0 & E_0 & \dots&  E_{N-3}\\
\vdots & \vdots &   & \ddots & \vdots\\
0 &0 & \dots & \dots & E_0 \end{array} \right)\,.
\ee
Computing the determinant and using Eq.\,\eqref{generating} we find 
\begin{align}\label{dimension adjoint}
\hat d_{\mu}(N)&=q^{-3(N-1)/2}\left(E_2E_1-E_3\right) \notag\\
&= \frac{\left[N \right]\left[N-1 \right]\left[N+1 \right]}{\left[3 \right]}\,.
\end{align}
Notice that due to property \eqref{classical}, in the classical ($k\to\infty$) limit, the dimensions \eqref{dimensions} and \eqref{dimension adjoint} reduce to the values obtained using the hook-length formula \eqref{UN dim}. Thus, displaying the pertinence of the terminology \emph{quantum dimension}.

The previous discussion suggests the possibility of finding a quantum version of the relative dimension. Indeed, the argument of Section \ref{proof} can be immediately generalized once we take into account some key features. For example, the analogue of Eq.\,\eqref{prod} for the WZW characters reads
\be \label{prod1}
\chi_\mu \chi_{\nu} =\sum_{\nu'} {\cal N}_{\mu,\nu}^{\;\;\nu'}\,  \chi_{\nu'} \;,
\ee
where the ${\cal N}_{\mu,\nu}^{\;\;\nu'}$ are the fusion coefficients of the model. In fact, these coefficients can be retrieved from the modular S-matrix \eqref{mod S} using the Verlinde formula \cite{Verlinde:1988sn}
\be
 {\cal N}_{\mu,\nu}^{\;\;\nu'}=\sum_\lambda\frac{ {\cal S}_{\nu\lambda} {\cal S}_{\mu\lambda} {\cal S}^{\lambda\nu'}}{ {\cal S}_{0\lambda}}\,.
\ee
Thus, we introduce the \emph{quantum relative dimension} 
\be\label{ReQD}
\widehat{\text{Dim}}[\mu,M;\nu, N]=\sum_{\nu'}  {\cal N}_{\mu,\nu}^{\;\;\nu'}\; \hat d_{\nu'}(M-N)\,.
\ee
Finally, using the Kac-Walton formula \cite{Walton:1990qs} it is easy to show that
\be \label{KW}
 \lim_{k\to\infty}{\cal N}_{\mu,\nu}^{\;\;\nu'}= g(\mu ;\nu,\nu ^{\prime }) \, .
\ee
Hence, the relative quantum dimension duly reduces to its classical counterpart in the large-$k$ limit
\be 
 \lim_{k\to\infty}\widehat{\text{Dim}}[\mu,M;\nu, N]={\text{Dim}}[\mu,M;\nu, N]\,.
\ee
Moreover, it is easy to show that 
\be
\widehat{\text{Dim}}[\mu,M;\emptyset, 0]=\hat d_{\mu}(M)\,.
\ee
Thus Eq.\,\eqref{ReQD} furnishes a consistent generalization of the notion of relative dimension to the context of 
affine Lie algebras and RCFTs. 
In future work, we shall explore physical applications of this formula as well as its relationship to affine branching rules. 

\section{Conclusions and outlook}

In this work we probed the structure of the branching graph of the unitary group using Schur transitions. We found that these transitions yield a new combinatorial expression for the relative dimensions of this graph. This formula, which is valid at any rank, is displayed in Eq.\,\eqref{relativedimension} and is one of the main results of this paper. In section \ref{proof} we establish the validity of this expression by providing a formal proof. As a first application of this formula we show that the Borodin-Olshanski identity can be succinctly derived. Indeed, it seems that large $N$ matrix model type techniques, such as the ones employed in this work, are proving to be highly effective at tackling questions in representation theory.

The form of equation \eqref{relativedimension} strongly suggests a quantum generalization. We define a notion of \emph{quantum relative dimension} in Eq.\,\eqref{ReQD} and subject it to the appropriate consistency tests. This new quantity finds its natural environment in the context of RCFTs and fractional statistics; where the already established notion of quantum dimension has proven to be of great physical importance.

\section*{Acknowledgments}

We would like to thank A. Borodin, P. Caputa, S. Das, V. Jejjala, R. de Mello Koch, G. Olshanski,  S. Ramgoolam and M. Walton for illuminating correspondence and conversations.  The research of PD is supported by the Natural Sciences and Engineering Research Council of Canada and the University of Lethbridge. GK would like to acknowledge The University of Johannesburg for financial assistance through the GES programme. The work of AVO is based upon research supported in part by the South African Research Chairs Initiative of the Department of Science and Technology and National Research Foundation.


\appendix
\section{Gelfand-Tsetlin patterns example}\label{example}

In this appendix, we list the Gelfand-Tsetlin patterns for a particular $U(N)$ irreducible representation so that the reader may gain more familiarity with these objects. Consider the vertex
\begin{equation}
(\mu,N) = \bigg(\,\raisebox{0.7ex}{\ytableausetup{boxsize=0.95em}\ydiagram[]{2,1}},3\,\bigg),
\end{equation}
whose signature is $(2,1,0)$. The number of valid Gelfand-Tsetlin patterns are eight in this case. They are:
\begin{eqnarray}\label{GT patterns}
&\left(
\begin{array}{ccccc}
2 &  & 1 &  & 0 \\
& 2 &  & 1 &  \\
&  & 2 &  &
\end{array}%
\right) &,\left(
\begin{array}{ccccc}
2 &  & 1 &  & 0 \\
& 2 &  & 1 &  \\
&  & 1 &  &
\end{array}%
\right) ,\left(
\begin{array}{ccccc}
2 &  & 1 &  & 0 \\
& 2 &  & 0 &  \\
&  & 2 &  &
\end{array}%
\right) ,\left(
\begin{array}{ccccc}
2 &  & 1 &  & 0 \\
& 2 &  & 0 &  \\
&  & 1 &  &
\end{array}%
\right) ,  \notag \\
&\left(
\begin{array}{ccccc}
2 &  & 1 &  & 0 \\
& 2 &  & 0 &  \\
&  & 0 &  &
\end{array}%
\right) ,&\left(
\begin{array}{ccccc}
2 &  & 1 &  & 0 \\
& 1 &  & 1 &  \\
&  & 1 &  &
\end{array}%
\right) ,\left(
\begin{array}{ccccc}
2 &  & 1 &  & 0 \\
& 1 &  & 0 &  \\
&  & 1 &  &
\end{array}%
\right) ,\left(
\begin{array}{ccccc}
2 &  & 1 &  & 0 \\
& 1 &  & 0 &  \\
&  & 0 &  &
\end{array}%
\right) .  \notag \\
&&
\end{eqnarray}%
Note that the rule is that in each level down, the numbers must be in between as the interlace condition dictates. Each row in a GT pattern is the signature of the irrep of the unitary group at the corresponding level, where that irrep has been subduced from the irrep corresponding to the level above. As described in section \ref{sec: GT and Y graphs}, each GT pattern is a path in $\mathbb{GT}$.

Lastly, the basis states in the carrier space of a $U(N)$ irrep are in one-to-one correspondence with the GT patterns. To illustrate this, consider again the above example. The dimension of this $U(3)$ irrep may be calculated from equation (\ref{UN dim}). The result is eight which is in one-to-one correspondence with the eight GT patterns in (\ref{GT patterns}). \newline

\section{Some relative dimensions computed by counting GT patterns}\label{examples}

In this appendix we give some extra examples of counting partial (or trapezoidal) GT patterns from a signature $[\mu,M]$ to signature $[\nu,N]$, which gives you the relative dimension $\text{Dim}[\mu,M;\nu,N]$. Remember that  we claim that 
\be
\text{Dim}[\mu,M;\nu,n]= \frac{1}{(m-n)!}\sum_{\nu ^{\prime }\vdash m-n} g(\mu ;\nu,\nu ^{\prime }) \text{dim}_{\nu' }f_{\nu'}[M-N], 
\ee
which is obtaind from (\ref{eq:finite}) after dividing by $\frac{\text{Dim}[\nu,N]}{\text{Dim}[\mu,M]}$. Let see some cases.\\
In the example
\begin{equation}
\text{Dim}\Big[\ytableausetup{boxsize=0.70em}\ydiagram[]{2},M; \emptyset,N\Big]=\frac{1}{2}(M-N)(M-N+1),
\end{equation}
we pass from signature $(\underbrace{2,0,\dots,0}_M)$ to signature $(\underbrace{0,0,\dots,0}_N)$ in $M-N$ steps.
An effective way of counting it is to consider the number of  ``1's'' that appear in a partial GT pattern. So with no ``1's''  we can write $M-N$ partial patterns. With one ``1'' we can write $M-N-1$, with two ``1's'' $M-N-2$ and so on. So the total number of partial GT patterns can be calculated as

\begin{equation}
\text{Dim}\Big[\ytableausetup{boxsize=0.70em}\ydiagram[]{2},M; \emptyset,N\Big]=\sum_{i=1}^{M-N} i=\frac{1}{2}(M-N)(M-N+1),
\end{equation}
as (\ref{relativedimension}) predicts.

Using the same kind of combinatorics one can calculate by ``brute force'' the following relative dimensions

\begin{eqnarray}
\text{Dim}\Big[\, \raisebox{0.6ex}{\ytableausetup{boxsize=0.70em}\ydiagram[]{1,1}},M; \ytableausetup{boxsize=0.70em}\ydiagram[]{1},N\Big]&=& M-N \nonumber \\
\text{Dim}\Big[\,\raisebox{0.6ex}{\ytableausetup{boxsize=0.70em}\ydiagram[]{1,1}},M; \emptyset,N\Big]&=&\frac{1}{2}(M-N)(M-N-1) \nonumber \\
\text{Dim}\Big[\, \raisebox{0.6ex}{\ytableausetup{boxsize=0.70em}\ydiagram[]{2,1}},M; \ytableausetup{boxsize=0.70em}\ydiagram[]{2},N\Big]&=& M-N \nonumber \\
\text{Dim}\Big[\,\raisebox{0.6ex}{\ytableausetup{boxsize=0.70em}\ydiagram[]{2,1}},M; \ytableausetup{boxsize=0.70em}\ydiagram[]{1,1},N\Big]&=& M-N \nonumber \\
\text{Dim}\Big[ \,\raisebox{0.6ex}{\ytableausetup{boxsize=0.70em}\ydiagram[]{2,1}},M; \ytableausetup{boxsize=0.70em}\ydiagram[]{1},N\Big]&=& (M-N)^2 \nonumber \\
\text{Dim}\Big[\,\raisebox{0.6ex}{\ytableausetup{boxsize=0.70em}\ydiagram[]{2,1}},M; \emptyset,N\Big]&=&\frac{1}{3}(M-N)(M-N-1)(M-N+1) \nonumber \\
\end{eqnarray}

 and match the prediction of (\ref{relativedimension}) .

The reader might be suspicious because the above examples are multiplicity free since LR numbers are always 1. Let us show in detail an 
example where this is not the case. We will compute the relative dimension
\be
\text{Dim}\Big[\, \raisebox{1.2ex}{\ytableausetup{boxsize=0.55em}\ydiagram[]{3,2,1}},M;  \raisebox{0.5ex}{\ytableausetup{boxsize=0.70em}\ydiagram[]{2,1}},N\Big].\label{eq:example}
\ee
Appying formula \eqref{relativedimension} we can see that 
\begin{eqnarray}
\text{Dim}\Big[\, \raisebox{1.2ex}{\ytableausetup{boxsize=0.55em}\ydiagram[]{3,2,1}},M;  \raisebox{0.5ex}{\ytableausetup{boxsize=0.70em}\ydiagram[]{2,1}},N\Big]&=&\text{Dim}_{\ytableausetup{boxsize=0.50em}\ydiagram[]{1,1,1}}[M-N]+\text{Dim}_{\ytableausetup{boxsize=0.50em}\ydiagram[]{3}}[M-N]+2\,\text{Dim}_{\ytableausetup{boxsize=0.50em}\ydiagram[]{2,1}}[M-N]\nonumber \\
&=& (M-N)^3,\label{exampleformula}
\end{eqnarray}
where the factor 4 in the third term is the product of the multiplicity and the dimension of the irrep, both being 2.\\
Let us compute directly (\ref{eq:example}). We should count all the paths in the graph that join the irrep with signature $(\underbrace{3,2,1,0,\dots,0}_M)$ with the irrep with signature  $(\underbrace{2,1,0,0,\dots,0}_N)$. In the following we will call
\be
[(3,2,1),M]\equiv(\underbrace{3,2,1,0,\dots,0}_M),\quad [(2,1),N]\equiv (\underbrace{2,1,0,0,\dots,0}_N).
\ee
 This is tantamount to counting the number of {\it partial} GT patterns we can write starting from $(3,2,1,0,\dots,0)$ and reaching $(2,1,0,\dots,0)$ in $M-N$ steps. We will solve a recursion relation for it. Since $\mathbb{GT}$ is multiplicity free, irreps at consecutive levels are either singled link or not linked. Using the interlazing rule of the graph we know that the second to last level, which is $N+1$, is connected to $[(2,1),N]$ via the following eight irreps:
\begin{eqnarray}
&&[(2,1), N+1],\quad [(2,2), N+1],\quad [(3,2), N+1],\quad [(3,3), N+1],\nonumber \\
&& [(3,1), N+1],\quad [(3,1,1), N+1],\quad [(2,2,1), N+1],\quad [(2,1,1), N+1],\quad 
\end{eqnarray}
so actually we can write 
\begin{small}
\begin{align}
\text{Dim}[(3,2,1),M;(2,1),N]&=\text{Dim}[(3,2,1),M;(2,1),N+1]+\text{Dim}[(3,2,1),M;(2,2),N+1] \nonumber \\
&=\text{Dim}[(3,2,1),M;(3,2),N+1]+\text{Dim}[(3,2,1),M;(3,2,1),N+1] \nonumber \\
&=\text{Dim}[(3,2,1),M;(3,1),N+1]+\text{Dim}[(3,2,1),M;(3,1,1),N+1] \nonumber \\
&=\text{Dim}[(3,2,1),M;(2,2,1),N+1]+\text{Dim}[(3,2,1),M;(2,1,1),N+1].\nonumber \\ \label{eq:rec} 
\end{align}
\end{small}
It turns out that seven of the eigth terms appearing in (\ref{eq:rec}) are easily computed. For example
\be
\text{Dim}[(3,2,1),M;(3,2),N+1]=M-N-1
\ee
is easy to verify since the last ``1''  in $(3,2,1)$ must go to ``0'' and it has $M-N-1$ locations. A similar reasoning can be used to see that
\be
 \text{Dim}[(3,2,1),M;(3,1,1),N+1]=\text{Dim}[(3,2,1),M;(2,2,1),N+1]=M-N-1.
\ee 
We can also see that 
\be
\text{Dim}[(3,2,1),M;(2,2),N+1]=(M-N-1)^2.
\ee
In this case note that the second ``2'' in $(3,2,1)$ cannot change, and ``3'' must go to ``2'' together with ``1'' going to ``0'' {\it independently}. Both have $M-N-1$ locations. The rest follow the same logic:
\be
 \text{Dim}[(3,2,1),M;(3,1),N+1]= \text{Dim}[(3,2,1),M;(2,1,1),N+1]=(M-N-1)^2.
\ee
We also know that 
\be
  \text{Dim}[(3,2,1),M;(3,2,1),N+1]=1,
\ee
that is, there is just one path that joins two irreps in the graph with the same Young Diagram independently of $M-N$.\\
Plugging all these results in (\ref{eq:rec}) we see that
\be 
\text{Dim}[(3,2,1),M;(2,1),N]=\text{Dim}[(3,2,1),M;(2,1),N+1] +1+3(M-N)(M-N-1).\label{eq:recursion}
\ee
This recurrence is easily solved. We start with
\be
\text{Dim}[(3,2,1),M;(2,1),M-1]=1,
\ee
since the graph is multiplicity free. Then we apply (\ref{eq:recursion}) to see that
\be 
\text{Dim}[(3,2,1),M;(2,1),M-2]=\text{Dim}[(3,2,1),M;(2,1),M-1]+1+3(M-(M-2)-1)(M-(M-2)).
\ee
We easily see that 
\be
\text{Dim}[(3,2,1),M;(2,1),M-j]=\sum_{i=1}^j[1+3i(i-1)].
\ee
Now, $M-j=N$ implies that $j=M-N$, so
\begin{eqnarray}
\text{Dim}[(3,2,1),M;(2,1),N]&=&\sum_{i=1}^{M-N}[1+3i(i-1)]\nonumber \\
&=&M-N+3\frac{(M-N-1)(M-N)(M-N+1)}{3}\nonumber \\
&=&(M-N)^3.
\end{eqnarray}
So, it exactly matches (\ref{exampleformula}).

\section{Schur functions and skew Schur functions}\label{schur f}
Let us write some definitions on Schur and skew Schur symmetric functions. More details can be found in \cite{Mac}.
Schur functions of $N$ variables furnish a basis of symmetric functions of those variables. They are labeled by Young diagrams and they can be defined as
\begin{equation}\label{Schurdef}
S_{\mu}(x_1,\dots,x_N)=\frac{1}{n!}\sum_{\sigma \in S_n}\chi_{\mu}(\sigma)p_{\sigma}(x_1,\dots,x_N),
\end{equation}
where $\mu$ is a Young diagram whose number of boxes $n$ determines the degree of the homogeneous polynomial $S_{\mu}(x_1,\dots,x_N)$. In (\ref{Schurdef}), $\chi_{\mu}(\sigma)$ is the character of the symmetric group corresponding to the irrep $\mu$ and evaluated at $\sigma\in S_n$, and $p_{\sigma}(x_1,\dots,x_N)$ are another basis of symmetric functions called {\it power sums} and defined as
\begin{equation}
p_{\sigma}(x_1,\dots,x_N)=\big(x_1^{\sigma_1}+\cdots+x_N^{\sigma_1}\big)\big(x_1^{\sigma_2}+\cdots+x_N^{\sigma_2}\big)\cdots
\big(x_1^{\sigma_r}+\cdots+x_N^{\sigma_r}\big),
\end{equation}
where $(\sigma_1,\dots,\sigma_r)$ is the cycle structure of $\sigma\in S_n$.\\
Schur polynomials labeled by $\mu$ are homogeneous of degree $m=|\mu|$. It is obvious that if we multiply two Schur polynomials {\it of the same variables} $S_{\nu}(x_1,\dots,x_N)  S_{\mu}(x_1,\dots,x_N)$, with $|\mu|=m$ and $|\nu|=n$, we obtain a homogeneous symmetric polynomial of degree $n+m$. This polynomial can of course be written in terms of Schur polynomials of $n+m$ degree. The coefficients that appear in this expansion are the Littlewood-Richardson numbers
\begin{equation}\label{Schurproduct}
S_{\nu}(x_1,\dots,x_N)  S_{\mu}(x_1,\dots,x_N)=\sum_{|\lambda|=m+n}g(\mu,\nu ;\lambda) S_{\lambda}(x_1,\dots,x_N).
\end{equation}
Schur functions can be defined in alternative ways, but the virtue of (\ref{Schurdef}) is that it makes explicit the {\it characteristic map} that relates class functions of the symmetric group (functions of $S_n$ that are invariant under the the change $\sigma\to g\sigma g^{-1}$) with symmetric functions. The symmetric function basis for characteristic maps is always $p_{\sigma}$. So definition (\ref{Schurdef}) can be expressed as ch: $\chi_{\mu}\mapsto S_{\mu}$.   

In the space of symmetric functions we can define an inner product which would assign a complex number to a pair of functions. This would act like and integration on the variables of the functions. Instead of defining such integral, since the inner product is a bilinear, it is customary to define it on every couple of elements  of a basis. The convention is to define the inner product on symmetric functions according to the characteristic map. Now, so since for characters we have the famous orthogonality relation
\begin{equation}
\frac{1}{n!}\sum_{\sigma\in S_n}\chi_{\mu}(\sigma)\chi_{\nu}(\sigma)=\delta_{\mu\nu},
\end{equation}  
the inner product of symmetric functions is usually defined as
\begin{equation}\label{innerprod}
\langle S_{\mu},S_{\nu}\rangle=\delta_{\mu\nu}.
\end{equation}
Sometimes it may be useful to define the inner product in other bases
\be
\langle h_\lambda, m_\mu \rangle=\delta_{\lambda \mu},\label{inner}
\ee
where $\{h_\lambda |\lambda\vdash n\}$ and $\{m_\mu |\mu\vdash n\}$ are the complete and the monomial basis of symmetric functions, respectively.

The definition of skew Schur functions $S_{\mu/\nu}$, which are homogeneous of degree $m-n$ of the variables in which $S_{\mu}$ and $S_{\nu}$ were originally defined, uses (\ref{Schurproduct}) and (\ref{innerprod}) and can be stated as
\be\label{skew}
\langle S_{\mu/\nu}, S_{\nu'} \rangle =\langle S_\mu, S_\nu S_{\nu'} \rangle,\quad \mu\vdash m,~\nu\vdash n, ~ \nu'\vdash m-n.
\ee

\section{Restricted Schur polynomials}\label{ResSchur}

An exactly orthogonal basis for the 1/2-BPS sector of free $\mathcal{N}=4$ super Yang-Mills theory with a $U(N)$ gauge group was found to be Schur polynomials $\chi_{R}(Z)$, labeled by irreducible representation (irrep) $R$ of the symmetric and unitary group \cite{Corley:2001zk}. Such a basis may be used to explore large $N$ non-planar limits of the theory. Furthermore, these operators have an interpretation in the dual string theory. When $R$ has long columns or long rows (each row or column having $O(N)$ boxes) then $\chi_{R}(Z)$ is dual to a system of giant gravitons in the $S^{5}$ or AdS$_{5}$ \cite{Corley:2001zk, McGreevy:2000cw, Grisaru:2000zn, Hashimoto:2000zp}.

This basis was then extended to 1/4-BPS sector of the gauge theory in the form of restricted Schur polynomials. Using two of the complex valued scalar fields, these operators are
\begin{eqnarray}
\label{eq:ResSchur}
\chi_{R(r,s)\alpha\beta}(Z,Y) = \frac{1}{n!m!}\sum\limits_{\sigma \in S_{n+m}}\chi_{R(r,s)\alpha\beta}(\sigma)\mathrm{Tr}(\sigma Z^{\otimes n}\otimes Y^{\otimes m}).
\end{eqnarray}
The restricted Schur may be thought of as a particular linear combination of all possible multi-matrix, multi-trace operators, where the sum is over permutations of the symmetric group. $R$ is an (irrep) of $S_{n+m}$ and is labeled by a Young diagram with $n+m$ boxes. Next, $(r,s)$ is an irrep of $S_{n}\times S_{m}$ that may be subduced from $R$, with $r$ and $s$ being Young diagrams with $n$ and $m$ boxes. The $\alpha$ and $\beta$ are multiplicity labels with which $(r,s)$ is subduced from $R$. Finally, $\chi_{R(r,s)\alpha\beta}(\sigma)$ is called the restricted character and is simply the trace of the matrix representing $\sigma$ in irrep $R$, but restricted to the block whose row index is labeled by $\alpha$ and whose column index is labeled by $\beta$. See \cite{de Mello Koch:2007uu} for further details. 

The exact two-point function of the restricted Schurs was computed in \cite{Bhattacharyya:2008rb} and found to be diagonal in the operator's labels. Also, the product of two Schur polynomials may be expanded in terms of the restricted Schurs. Letting $r$ and $s$ be Young diagrams with $n$ and $m$ boxes, \cite{Bhattacharyya:2008xy} finds
\begin{eqnarray}
\label{eq:productrule}
\chi_{r}(Z)\chi_{s}(Y) = \frac{n!m!}{(n+m)!d_{r}d_{s}}\sum\limits_{T,t,u,\gamma,\rho} d_{T} \chi_{T(t,u)\gamma\rho}(Z,Y)
\end{eqnarray}
The expansion coefficients involve simple group theoretic factors such as the dimension of an irrep. When $R$ has $O(N)$ boxes in each column or row, and $n \gg m$, the interpretation of (\ref{eq:ResSchur}) in the string theory is that of a system of giant gravitons with $m$ strings attached \cite{de Mello Koch:2007uu}. Amongst the other bases found for the 1/4-BPS sector it has been argued that restricted Schur polynomials is the most natural basis for studying open string dynamics of their dual D-brane states \cite{Bhattacharyya:2008xy}. To this end, the spectral problem of restricted Schurs and its dual system has been extensively studied in \cite{Koch:2010gp, Koch:2011hb, Carlson:2011hy,deMelloKoch:2012ck, deMelloKoch:2011ci, deMelloKoch:2012sv}.

Generalizations of restricted Schur polynomials to fermion fields, gauge fields and three complex scalar fields have been studied in \cite{Koch:2011jk, Koch:2012sf, deMelloKoch:2011vn}. Generalizations of restricted Schurs to an $SO(N)$ gauge group have been studied in \cite{G1, G2}.




\begin{thebibliography}{99}


\bibitem{BO} A. Borodin, G. Olshanski,
{\it The Young bouquet and its boundary,}
Mosc. Math. J. \textbf{13} (2013) 193-232 [arXiv:1110.4458].

\bibitem{Ramgoolam:2008yr} S.~Ramgoolam,
{\it Schur-Weyl duality as an instrument of Gauge-String duality,}
AIP Conf.\ Proc.\ \textbf{1031} (2008) 255 [arXiv:0804.2764 [hep-th]].



\bibitem{Diaz:2015tda}
  P.~Diaz, H.~Lin and A.~Veliz-Osorio,
  ``Graph duality as an instrument of Gauge-String correspondence,''
  arXiv:1505.04837 [hep-th].



\bibitem{Wen:1990se}
  X.~G.~Wen,
  ``Chiral Luttinger Liquid and the Edge Excitations in the Fractional Quantum Hall States,''
  Phys.\ Rev.\ B {\bf 41} (1990) 12838.
  doi:10.1103/PhysRevB.41.12838

\bibitem{Moore:1991ks}
  G.~W.~Moore and N.~Read,
  ``Nonabelions in the fractional quantum Hall effect,''
  Nucl.\ Phys.\ B {\bf 360} (1991) 362.
  doi:10.1016/0550-3213(91)90407-O

\bibitem{Moore:1989vd}
  G.~W.~Moore and N.~Seiberg,
  ``Lectures On Rcft,''
  RU-89-32, YCTP-P13-89, C89-08-14.





\bibitem{Kitaev:2005dm}
  A.~Kitaev and J.~Preskill,
  ``Topological entanglement entropy,''
  Phys.\ Rev.\ Lett.\  {\bf 96} (2006) 110404
  [hep-th/0510092].

\bibitem{He:2014mwa}
  S.~He, T.~Numasawa, T.~Takayanagi and K.~Watanabe,
  ``Quantum dimension as entanglement entropy in two dimensional conformal field theories,''
  Phys.\ Rev.\ D {\bf 90} (2014) no.4,  041701
  [arXiv:1403.0702 [hep-th]].

\bibitem{Caputa:2015tua}
  P.~Caputa and A.~Veliz-Osorio,
  ``Entanglement constant for conformal families,''
  Phys.\ Rev.\ D {\bf 92} (2015) no.6,  065010
  [arXiv:1507.00582 [hep-th]].

\bibitem{Caputa:2016tgt}
  P.~Caputa, T.~Numasawa and A.~Veliz-Osorio,
  ``Scrambling without chaos in RCFT,''
  arXiv:1602.06542 [hep-th].


\bibitem{Gu:2016hoy}
  Y.~Gu and X.~L.~Qi,
  ``Fractional Statistics and the Butterfly Effect,''
  arXiv:1602.06543 [hep-th].

\bibitem{BO2012} A. Borodin, G. Olshanski,
{\it Markov processes on the path space of the Gelfand-Tsetlin graph and on its boundary,}
J. Funct. Anal. \textbf{263} (2012) 248-303 [arXiv:1009.2029].

\bibitem{Maldacena:1997re} J. M.~Maldacena,
{\it The Large-N limit of superconformal field theories and supergravity,}
Adv. Theor. Math. Phys. \textbf{2} (1998) 231 [Int. J. Theor. Phys. 38 1113
(1999)] [hep-th/9711200].


\bibitem{Lin:2004nb} H.~Lin, O.~Lunin and J.~M.~Maldacena,
{\it Bubbling AdS space and 1/2 BPS geometries,}
JHEP \textbf{0410} (2004) 025 [hep-th/0409174].


\bibitem{Koch:2008ah} R.~d.~M.~Koch, 
{\it Geometries from Young Diagrams,}
JHEP \textbf{0811} (2008) 061 [arXiv:0806.0685 [hep-th]].



\bibitem{Bhattacharyya:2008xy} R.~ Bhattacharyya, R.~de Mello Koch, and M.~Stephanou,
{\it Exact Multi-Restricted Schur Polynomial Correlators,}
JHEP \textbf{0806} (2008) 101 [arXiv:0805.3025 [hep-th]].

\bibitem{Bhattacharyya:2008rb} R.~Bhattacharyya, S.~Collins and
R.~d.~M.~Koch, {\it Exact Multi-Matrix Correlators,}
JHEP \textbf{0803} (2008) 044 [arXiv:0801.2061 [hep-th]].


\bibitem{Mac} I.~G.~MacDonald, ``Symmetric functions and Hall polynomials'', Oxford University Press,
1995.



\bibitem{Witten:1988hf}
  E.~Witten,
  ``Quantum Field Theory and the Jones Polynomial,''
  Commun.\ Math.\ Phys.\  {\bf 121} (1989) 351.


\bibitem{Dong:2008ft}
  S.~Dong, E.~Fradkin, R.~G.~Leigh and S.~Nowling,
  ``Topological Entanglement Entropy in Chern-Simons Theories and Quantum Hall Fluids,''
  JHEP {\bf 0805} (2008) 016
 [arXiv:0802.3231 [hep-th]].


\bibitem{Verlinde:1988sn}
  E.~P.~Verlinde,
  ``Fusion Rules and Modular Transformations in 2D Conformal Field Theory,''
  Nucl.\ Phys.\ B {\bf 300} (1988) 360.


\bibitem{Walton:1990qs}
  M.~A.~Walton,
  ``Algorithm for {WZW} Fusion Rules: A Proof,''
  Phys.\ Lett.\ B {\bf 241} (1990) 365
   Erratum: [Phys.\ Lett.\ B {\bf 244} (1990) 580].


\bibitem{Corley:2001zk} S.~Corley, A.~Jevicki and S.~Ramgoolam,
{\it Exact correlators of giant gravitons from dual N=4 SYM theory,}
Adv.\ Theor.\ Math.\ Phys.\ \textbf{5} (2002) 809 [hep-th/0111222].


\bibitem{McGreevy:2000cw} J.~McGreevy, L.~Susskind and N.~Toumbas,
{\it Invasion of the giant gravitons from anti-de Sitter space,}
JHEP \textbf{0006} (2000) 008 [hep-th/0003075].

\bibitem{Grisaru:2000zn} M. T.~Grisaru, R. C.~Myers and O.~Tafjord,
{\it SUSY and Goliath,}
JHEP \textbf{0008} (2000) 040 [hep-th/0008015].

\bibitem{Hashimoto:2000zp} A. Hashimoto, S. Hirano and N. Itzhaki,
{\it Large branes in AdS and their field theory dual,}
JHEP \textbf{0008} (2000) 051 [hep-th/0008016].

\bibitem{de Mello Koch:2007uu} R.~de Mello Koch, J.~Smolic and M.~Smolic,
{\it Giant Gravitons - with Strings Attached(I),}
JHEP \textbf{0706}(2007) 074 [hep-th/0701066].



\bibitem{Koch:2010gp} 
  R.~d.~M.~Koch, G.~Mashile and N.~Park,
  {\it Emergent Threebrane Lattices,}
  Phys.\ Rev.\ D {\bf 81}, 106009 (2010)
  [arXiv:1004.1108 [hep-th]]. $\bullet$   V.~De Comarmond, R.~de Mello Koch and K.~Jefferies,
  {\it Surprisingly Simple Spectra,}
  JHEP {\bf 1102}, 006 (2011)
  [arXiv:1012.3884 [hep-th]]. 

\bibitem{Koch:2011hb} R.~d.~M.~Koch, M.~Dessein, D.~Giataganas and
C.~Mathwin, {\it Giant Graviton Oscillators,}
JHEP \textbf{1110} (2011) 009 [arXiv:1108.2761 [hep-th]].

\bibitem{Carlson:2011hy} W.~Carlson, R.~d.~M.~Koch and H.~Lin,
{\it Nonplanar Integrability,}
JHEP \textbf{1103} (2011) 105 [arXiv:1101.5404 [hep-th]].


\bibitem{deMelloKoch:2012ck} R.~de Mello Koch and S.~Ramgoolam,
{\it A double coset ansatz for integrability in AdS/CFT,}
JHEP \textbf{1206} (2012) 083 [arXiv:1204.2153 [hep-th]].


\bibitem{deMelloKoch:2011ci} R.~de Mello Koch, G.~Kemp and S.~Smith,
{\it From Large N Nonplanar Anomalous Dimensions to Open Spring Theory,}
Phys.\ Lett.\ B \textbf{711} (2012) 398 [arXiv:1111.1058 [hep-th]].

  \bibitem{deMelloKoch:2012sv}
  R.~de Mello Koch, G.~Kemp, B.~A.~E.~Mohammed and S.~Smith,
  {\it Nonplanar integrability at two loops,}
  JHEP {\bf 1210}, 144 (2012)
  [arXiv:1206.0813 [hep-th]].

\bibitem{Balasubramanian:2001nh} V.~Balasubramanian, M.~Berkooz, A.~Naqvi and M.~J.~Strassler,
{\it Giant gravitons in conformal field theory,}
JHEP {\bf 0204} (2002) 034 [hep-th/0107119].




\bibitem{Koch:2011jk} 
  R.~d.~M.~Koch, B.~A.~E.~Mohammed and S.~Smith,
  {\it Nonplanar Integrability: Beyond the SU(2) Sector,}
  Int.\ J.\ Mod.\ Phys.\ A {\bf 26}, 4553 (2011)
  [arXiv:1106.2483 [hep-th]].

\bibitem{Koch:2012sf} R.~de Mello Koch, P.~Diaz and N.~Nokwara,
{\it Restricted Schur Polynomials for Fermions and integrability in the su(2$|$3) sector,}
JHEP \textbf{1303} (2013) 173 [arXiv:1212.5935 [hep-th]].

\bibitem{deMelloKoch:2011vn} R.~de Mello Koch, P.~Diaz and H.~Soltanpanahi,
{\it Non-planar Anomalous Dimensions in the sl(2) Sector,}
Phys.\ Lett.\ B \textbf{713} (2012) 509 [arXiv:1111.6385 [hep-th]].


\bibitem{G1} G.~ Kemp,
{\it SO(N) restricted Schur polynomials,}
J. Math. Phys. {\bf 56} (2015) 022302 [arXiv:1405.7017 [hep-th]].

\bibitem{G2} G.~ Kemp,
{\it Restricted Schurs and correlators for SO(N) and Sp(N),}
JHEP {\bf 1408} (2014) 137 [arXiv:1406.3854 [hep-th]].




\end{thebibliography}
\end{document}